\documentclass{jpp}
\usepackage{array}
\usepackage{graphicx}
\usepackage{multirow}
\usepackage{hyperref}
\usepackage[utf8]{inputenc}
\usepackage[T1]{fontenc}
\usepackage{amsmath}
\usepackage{tikz}
\usepackage{graphicx} %
\usepackage{amsmath, amsfonts}
\usepackage{bm}
\DeclareMathOperator*{\argmin}{argmin}
\DeclareMathOperator*{\argmax}{argmax}
\DeclareMathSymbol{\sm}{\mathbin}{AMSa}{"39}
\newcommand{\etabar}{\overline{\eta}}

\usepackage{tikz}
\usepackage{url}
\usepackage{diagbox}

\shorttitle{A comprehensive exploration of quasisymmetric stellarators and their coil sets}
\shortauthor{A. Giuliani, E. Rodr\'{i}guez, M. Spivak}

\title{A comprehensive exploration of quasisymmetric stellarators and their coil sets}

\author{Andrew Giuliani\aff{1}
  \corresp{\email{agiuliani@flatironinstitute.org}},
  Eduardo Rodr\'{i}guez\aff{2}
 \and Marina Spivak\aff{1}}

\affiliation{\aff{1}Center for Computational Mathematics, Flatiron Institute, 162 Fifth Avenue, New York, 10128, USA
\aff{2} Max Planck Institute of Plasma Physics at Greifswald, Wendelsteinstraße 1, 17491 Greifswald, Germany
}

\begin{document}

\maketitle

\begin{abstract}
We augment the `QUAsi-symmetric Stellarator Repository' (QUASR) to include vacuum field stellarators with quasihelical symmetry using a globalized optimization workflow.  The database now has almost 370,000 quasisaxisymmetry and quasihelically symmetric devices along with coil sets, optimized for a variety of aspect ratios, rotational transforms, and discrete rotational symmetries.
This paper outlines a couple of ways to explore and characterize the data set.  We plot devices on a near-axis quasisymmetry landscape, revealing close correspondence to this predicted landscape. 
We also use principal component analysis to reduce the dimensionality of the data so that it can easily be visualized in two or three dimensions.
Principal component analysis also gives a mechanism to compare the new devices here to previously published ones in the literature.
We are able to characterize the structure of the data, observe clusters, and visualize the progression of devices in these clusters.
The topology of the data is governed by the interplay of the design constraints and valleys of the quasisymmetry objective.
These techniques reveal that the data has structure, and that typically one, two or three principal components are sufficient to characterize it.
The latest version of QUASR is archived at \url{https://zenodo.org/doi/10.5281/zenodo.10050655} and can be explored online at \url{quasr.flatironinstitute.org}.
\end{abstract}


\section{Introduction}

In recent years, the stellarator community has developed novel numerical optimization algorithms to quickly design stellarators that are quasisymmetric.   Quasisymmetry (QS) is a property of the magnetic field, where the field strength on a given magnetic surface can be written as a linear combination of the Boozer angles \citep{boozer1981plasma} on the surface \citep{helander2014theory}.
It can be shown that such fields confine guiding center orbits \citep{boozer1981transport, NUHRENBERG1988113, rodriguez2020necessary,burby2020some}.
Gradient-based optimization algorithms have facilitated the design of magnetic fields with the QS property \citep{lp}, as well as the electromagnetic coils that generate them \citep{wechsung2022precise}.

There is interest in the stellarator community to explore the landscape of QS stellarators. Examples of such surveys can be found in the literature, including a set of four stellarators with highly precise volumetric quasisymmetry in \cite{lp}, a similarly generated data set of more than 150 devices by \cite{buller_family_2024}, and one of over 30 generated exploiting adjoint-based algorithms by \cite{nies2024exploration}.
An even larger data set of 500,000 stellarators was presented in \cite{Landreman_2022}, which leverages a model of quasisymmetric stellarators near their magnetic axes \citep{garren1991existence,landreman2019direct, Landreman_Sengupta_2019,rodriguez2023constructing}.
The devices in the data sets mentioned above all were released without the coil sets that generate them.
For data sets that include their coils we must turn to the examples of over 40 stellarator coil sets studied by \cite{kappel_magnetic_2024} or the work of \cite{jorge2024simplifiedflexiblecoilsstellarators}, which includes a set of over 10 devices with simple and flexible coils.

A globalized coil design workflow was recently described in \cite{quasr1}, and was used to generate a large database of quasiaxisymmetric (QA) stellarators and their associated coil sets.  The database is called QUASR for `QUAsi-symmetric Stellarator Repository' and can be explored online at \url{quasr.flatironinstitute.org}.  In this work, we extend these algorithms to search for quasihelically (QH) symmetric devices and augment QUASR so that it now contains around 370,000 devices with associated coil sets.  
Using this large data set, we explore the landscape of stellarators that are generated by modular electromagnetic coils.
To start, we present a few devices in QUASR and analyze their physics properties.
We then apply a variety of techniques to visualize and analyze this large data set. The first uses the near-axis model and its landscape of excellent and poor quasisymmetry \citep{rodriguez2022phases,rodriguez2023constructing}, which we call the near-axis quasisymmetry landscape, to frame some configurations in the data set.  
The second, and as an alternative, we employ a dimensionality reduction approach called principal component analysis (PCA) \citep{jolliffe1990principal}.  This is a well-known technique frequently used in data science for dimensionality reduction and visualization \citep{bishop2006pattern}.   It enables the practitioner to project the high-dimensional data set onto a linear manifold, e.g., a two-dimensional plane or three-dimensional volume, to visualize relationships and possible formation of clusters. 
We compute two quantitative scores that measure how faithful the lower-dimensional representation is to the higher-dimensional data.

To summarize, the main goals of this paper are twofold: (1) we extend QUASR to include both QA and QH stellarators (Sections \ref{sec:workflow}, \ref{sec:notable}), and (2) we use a couple of visualization techniques to explore the data set and gain insight on the devices in it (Sections \ref{sec:nae_landscape}, \ref{sec:PCA}).

\section{The coil design workflow} \label{sec:workflow}
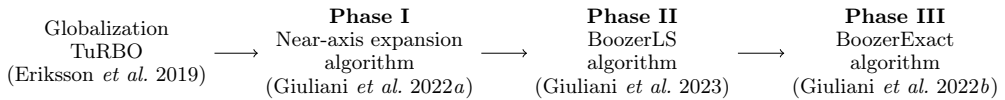
\begin{figure}
\centering
\resizebox{\textwidth}{!}{
\begin{tikzpicture}
    \node [align=center](A) at (0,0) {Globalization\\TuRBO\\ \citep{turbo}};
    \node  [align=center](B) at (4,0) {\textbf{Phase I}\\Near-axis expansion\\ algorithm \\
 \citep{nae}};
    \node [align=center](C) at (8,0) {\textbf{Phase II}\\BoozerLS\\algorithm \\ \citep{surfaceopt2}};
    \node [align=center](D) at (12,0) {\textbf{Phase III}\\BoozerExact\\algorithm \\ \citep{surfaceopt1}};
    \draw [->]  (A) edge (B) (B) edge (C) (C) edge (D) ; 
\end{tikzpicture}
}
    \caption{Globalized coil design workflow.}
    \label{fig:workflow}
\end{figure}

Our goal is to find coil geometries and currents that produce magnetic fields with the QS property (either QA or QH).
To achieve this, we use a globalized coil design workflow that comprises three phases, wrapped in a globalization algorithm.    
Since this workflow and its constituent algorithms have been described before \citep{turbo, quasr1, nae, surfaceopt2, surfaceopt1}, we only provide a brief overview here, summarized in Figure \ref{fig:workflow}.  The first phase searches for stellarator coils that are quasisymmetric near their axis.  This algorithm is robust, but may only find devices with nested flux surfaces on a small region in the neighborhood of the magnetic axis. These devices are then provided as an initial guess to Phases II and III which heal generalized chaos, optimize for nested flux surfaces, and polish for precise quasisymmetry.  The three phases are wrapped in a globalization algorithm \citep{turbo}, ensuring that the objective landscape is sufficiently well-explored.

To proceed with the volumetric refinement of quasisymmetry of Phases II and III, we must define some measure $f_\mathrm{QS}$ for the field generated by the electromagnetic coils to minimize. To do so, we must take into account the two flavors of quasisymmetry we are interested in: quasiaxisymmetry and quasihelical symmetry. The approaches for both symmetries are very similar, so we explain the process for quasiaxisymmetry, and then show how it can be extended to quasihelical symmetry.
\par
Quasiaxisymmetry is characterized by $\theta$ isolines of $B(\varphi, \theta) = \|\mathbf B(\bm \Sigma_s(\varphi, \theta)) \|$, on magnetic flux surfaces $\bm \Sigma_s$ parametrized in Boozer coordinates $(\varphi, \theta)$ \citep{boozer1981plasma}.
The field's deviation from exact quasiaxisymmetry on the magnetic surface can be computed by evaluating:
\begin{equation}\label{eq:qa1}
f_{QS} = \frac{\int^{1/n_{\text{fp}}}_{0}\int^1_0(B(\varphi, \theta) - B_{\text{QA}}(\theta) )^2 \biggl\| 
\frac{\partial \bm\Sigma_s}{\partial \varphi} \times \frac{\partial \bm\Sigma_s}{\partial \theta} \biggr\| ~d\theta ~d\varphi}{\int^{1/n_{\text{fp}}}_{0}\int^1_0 B_{\text{QA}}(\varphi, \theta)^2 \biggl\| 
\frac{\partial \bm\Sigma_s}{\partial \varphi} \times \frac{\partial \bm\Sigma_s}{\partial \theta} \biggr\|~d\theta ~d\varphi},
\end{equation}
where
\begin{equation}\label{eq:qa2}
B_{\text{QA}}(\theta) = \frac{\int^{1/n_{\text{fp}}}_{0} B(\varphi,\theta)~\| 
\frac{\partial \bm\Sigma_s}{\partial \varphi} \times \frac{\partial \bm\Sigma_s}{\partial \theta} \|~d\varphi}{\int^{1/n_{\text{fp}}}_{0} ~ \| 
\frac{\partial \bm\Sigma_s}{\partial \varphi} \times \frac{\partial \bm\Sigma_s}{\partial \theta} \|~d\varphi}
\end{equation}
is a least squares projection of $B(\varphi, \theta)$ onto the space of functions that do not depend on $\varphi$.
These integrals can easily be evaluated using the quadrature points on the tensor product grid $(\varphi_i, \theta_j) \in [0, 1/n_{\text{fp}}) \times [0, 1)$, where $\varphi_i = i/(N_{\varphi}n_{\text{fp}})$ and $\theta_j = j/N_{\theta}$ for $i=0, \hdots, N_{\varphi}-1$, $j = 0, \hdots, N_{\theta}-1$, and $N_{\varphi}, N_{\theta}$ are the number of quadrature points in the toroidal and poloidal directions, respectively.  These points correspond to the periodic trapezoidal rule, which is spectrally accurate \citep{trefethen2014exponentially}. This measure of quasisymmetry is similar to the commonly used sum over the symmetry breaking components of $B$ (defined as $f_B$ in \cite{rodriguez2022measures}).
To extend $f_\mathrm{QS}$ to the quasihelically symmetric case, we note that the difference is in the isolines of $B(\varphi, \theta)$, which now correspond to lines of constant $\theta- Nn_{\text{fp}}\varphi$, for $N\in\mathbb{Z}_{\neq0}$.
Most devices of interest have isolines with a slope exactly equal to $\pm n_{\text{fp}}$ \citep{rodriguez2022phases,rodriguez2023constructing,Landreman_2022}.
The same formulas and quadrature points used for QA can also be used in the case of QH by applying equations \eqref{eq:qa1}-\eqref{eq:qa2} in a rotated frame. %

To generate multiple devices through the phases described above, we scan over a number of physics and engineering target characteristics.
In terms of physics properties, we target various values of mean rotational transform $\iota$, aspect ratio, and total coil length, while keeping the major radius of 1 meter.  
The engineering characteristics on the coils impose limits on their curvature and mean squared curvature, which may not exceed $5~\text{m}^{-1}$ and $5~\text{m}^{-2}$, respectively.  The coils are also designed to have uniform incremental arclength. We refer to these as \textit{engineering constraints}.
The goal of these parameter scans is to explore the space of quasisymmetric stellarators that can be generated by coil sets, and observe possible trade-offs between competing physics and engineering characteristics.
Whenever `quality of quasisymmetry' for a device is provided in the manuscript, it is calculated by taking the square root of the average of \eqref{eq:qa1} on a number of surfaces of the magnetic field, which we may refer to as $\Bar{f}_\mathrm{QS}$.
Note that by construction, the surfaces on which we optimize for quasisymmetry in the final stage of the workflow are tangent the magnetic field at a fixed number of quadrature points on the surface.  As a result, $\mathbf B \cdot \mathbf n$ error is numerically zero when queried at those same quadrature points.  This is notably different from the approach in \citep{Jorge_2023}, where the $\mathbf B \cdot \mathbf n$ error is explicitly minimized.

The stellarator design problem amounts to a constrained optimization problem, where the ODE (phase 1) or PDE (phase 2, 3) constraint is satisfied to numerical precision at each iteration of the optimization loop.
This was not necessarily the case for the physics and engineering constraints.
However, all the target physics and engineering constraints were satisfied to a precision of $0.1\%$ by using a penalty method.  
After a fixed budget of iterations of the minimization algorithm, we increased by a factor of 10 the penalty weights associated to the constraints that were violated by more than this accuracy threshold. Then, with these new set of weights, the minimization algorithm is restarted.  This ensures that these constraints are satisfied to the desired accuracy at the end of the optimization workflow.
After a fixed number of restarts, devices that do not satisfy the engineering and physics constraints to $0.1\%$ accuracy, even after the penalty reweighting, are discarded.

\section{A few QH designs in QUASR} \label{sec:notable}
\begin{figure}
\centering
\includegraphics[width=\textwidth]{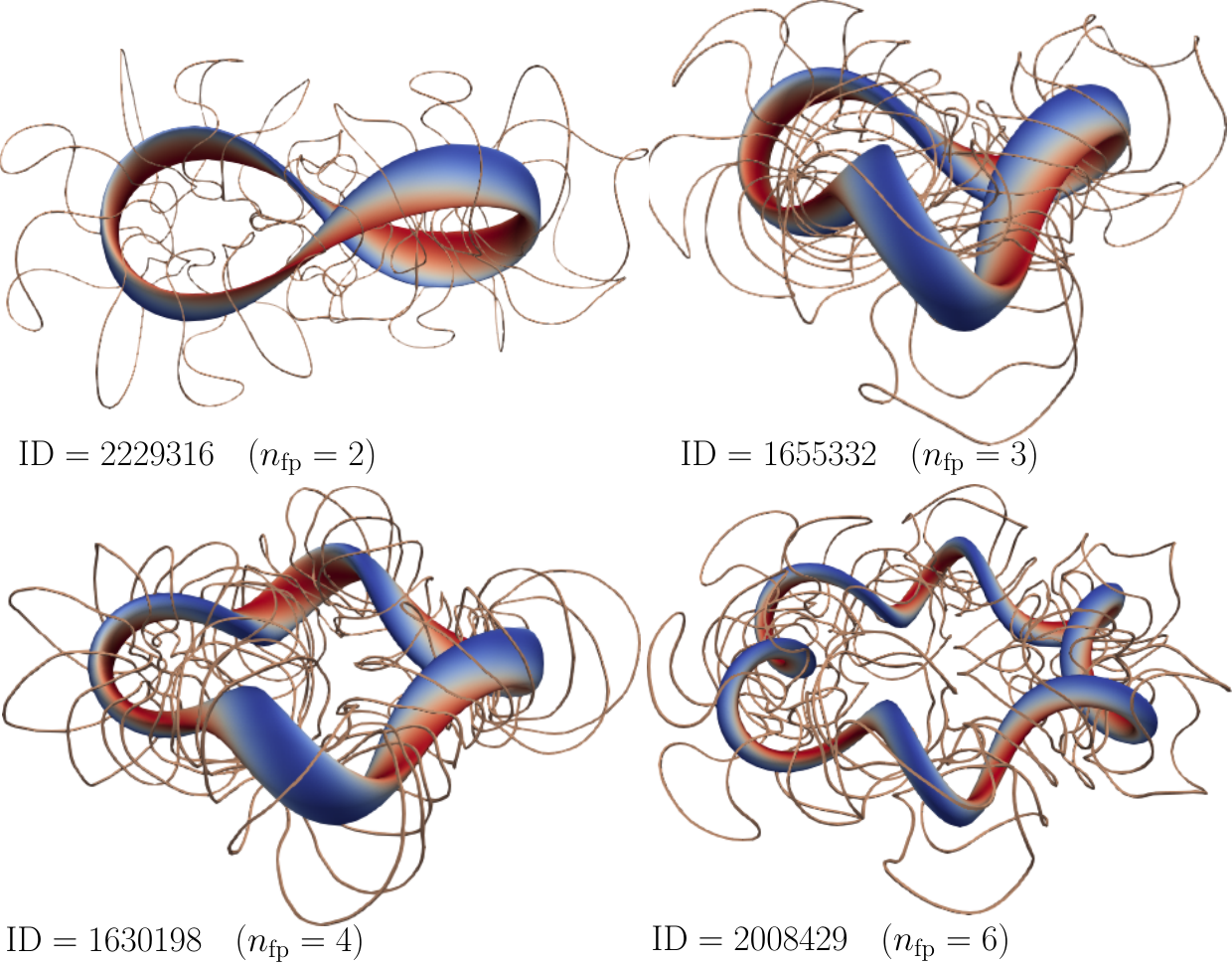}
    \caption{Notable QH devices and their coil sets in the QUASR.}
    \label{fig:devices}
\end{figure}

In this section, we present a few devices in the data set, summarized in Table \ref{tab:notable_devices} and plotted in Figure \ref{fig:devices}.  
The table also includes a link to a device manifest page, where additional physics and engineering information is provided.
We estimate the collisionless guiding-center \citep{blank2004guiding} losses of alpha particles in vacuum magnetic generated the coils of these devices after scaling them up to the ARIES-CS \citep{najmabadi2008aries} minor radius and spawning 5,000 alpha particles isotropically at half-radius in SIMSOPT \citep{landreman2021simsopt}, then tracing until $0.2$ seconds.   
All devices have favorable particle loss estimates despite deviations from QS, most likely owing to the narrow orbit widths in QH devices \citep{lp, buller_family_2024, paul2022energetic}. 
Except for the two field period device, all presented devices have reasonable elongation of flux surfaces in the $RZ$ plane. Note that this measure of shaping (especially when large) can be somewhat misleading in devices with a large helical excursion of the magnetic axis (see Section \ref{sec:qh_nfp4_iota2p3} for additional discussion on this). In such cases it is more instructive to measure the elongation in the plane perpendicular to the magnetic axis; for the $n_{\text{fp}}=2$ device, the maximum elongation in such plane reduces to $6.85$, which is still significant.This quantity was numerically computed by fitting a near-axis model to the field (Section \ref{sec:fitting}), and providing the parameters to \texttt{pyQSC} \citep{pyqsc}.

The first device is a two field period QH field with a `crossing point' with little space for coils, for which coils may nevertheless be found. Although it has the highest particle losses compared to the other devices presented here, the device is reminiscent of the original figure-eight stellarator design \citep{spitzer1958stellarator}.

The $n_{\text{fp}}=4$ device is quite similar to the Landreman-Paul precise QH configuration \citep{lp}, but was discovered here independently with coils. 
Coils for the precise QH device have been presented in \citep{Wiedman_Buller_Landreman_2024}, and found using a stage II algorithm \citep{Zhu_2018}.
Both device $\text{ID}=1630198$ and the Wiedman et al. device use 5 coils per half-period and a total coil length of approximately 104 meters, when scaled to a major radius of 1 meter.
All the coil sets in QUASR were designed to have a minimum coil-surface and coil-coil distance of $0.1 \text{m}$, and maximum coil curvature and mean squared curvature of $5 \text{m}^{-1}$ and $35 \text{m}^{-2}$, respectively.  
These were arbitrary choices, and \citep{Wiedman_Buller_Landreman_2024} used different thresholds (some more strict and some less).
Both designs though have remarkable estimated alpha particle losses ($0.00 \%$) at reactor scale.
Modular coils that wrap poloidally around the magnetic surfaces are just one coil topology and one might conceive of designs for the Landreman-Paul precise QH device with dipole \citep{Kaptanoglu_2025} or helical \citep{KAPTANOGLU2024116504} coils.

The $n_\mathrm{fp}=6$ device shows an example of a configuration with a high number of field periods, which are configurations that have proved challenging to find and provide a set of coils for \citep{Landreman_2022}.
\par
A note must be made on the coils presented for these configurations here. The filamentary coils sometimes appear quite long, which might not make the coil sets particularly practical. However, having such coils remains useful at least as an initial starting point for further refinement.

    \begin{table}
        \centering
        \begin{tabular}{cccccccc}
            ID & $n_{\text{fp}}$   & mean $\iota$ & aspect ratio  & max. elongation  & losses (\%) & quality of QS & device manifest\\
          2229316 &  2 & 0.9 & 6  & 8.75 & 3.26 & 4.03e-04 & \href{https://quasr.flatironinstitute.org/model/2229316}{[link]}\\
          1655332 &  3& 1.2 & 6  & 3.66 & 0.00 & 4.22e-04 & \href{https://quasr.flatironinstitute.org/model/1655332}{[link]}\\
          1630198 &  4& 1.2 & 8  & 3.62 &0.00 & 7.54e-04 & \href{https://quasr.flatironinstitute.org/model/1630198}{[link]}\\
          1641080 &  5& 1.5 & 8  & 2.16 & 0.00 & 2.26e-03 & \href{https://quasr.flatironinstitute.org/model/1641080}{[link]}\\
         2008429 &  6& 2.5 & 12 & 2.31 & 0.14 & 8.09e-04 & \href{https://quasr.flatironinstitute.org/model/2008429}{[link]}\\
        \end{tabular}
        \caption{Characteristics of some QH devices from the QUASR data set.  The maximum elongation is computed in the $RZ$ plane.  
        The final column of the table contains links to a device manifest page that contain more physics and engineering information about each design.
        }
        \label{tab:notable_devices}
    \end{table}

\section{The near-axis quasisymmetry landscape} \label{sec:nae_landscape}
The landscape of quasisymmetry quality in terms of the coil degrees of freedom is complex and high-dimensional, which can be difficult to visualize.
Using a near-axis framework, though, this landscape gains significant structure when parametrized in terms of the magnetic axis shape, owing to the key role played by topological properties of these \citep{rodriguez2022phases}. A representation of a representative subset of these axes described only by a few parameters (and thus lower dimensionality) can then provide a simple way to assess and visualize QS fields \citep{rodriguez2023constructing}.

\begin{figure}
    \centering
    \includegraphics[width=\textwidth]{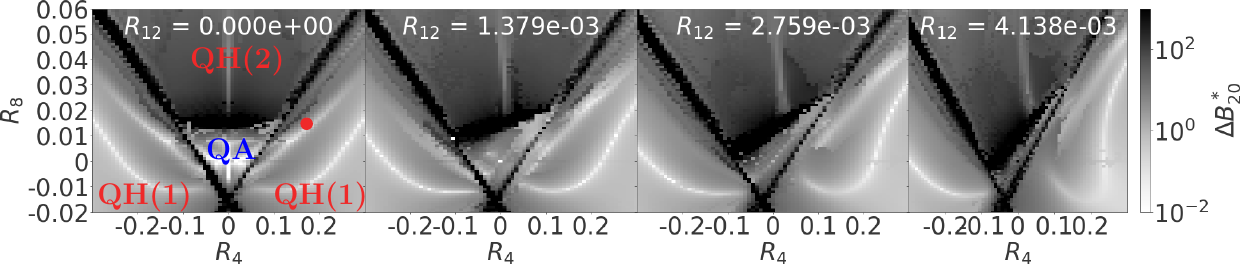}
    \includegraphics[width=\textwidth]{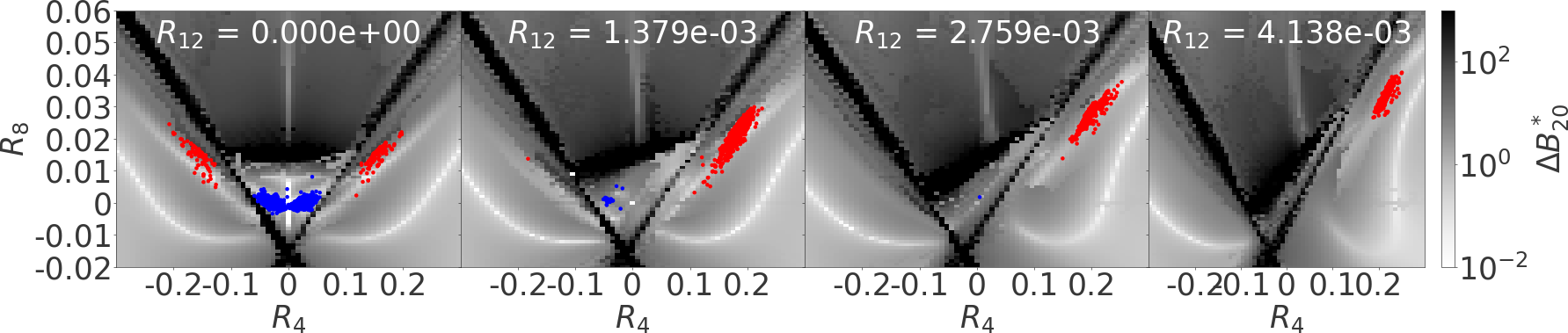}
    \includegraphics[width=\textwidth]{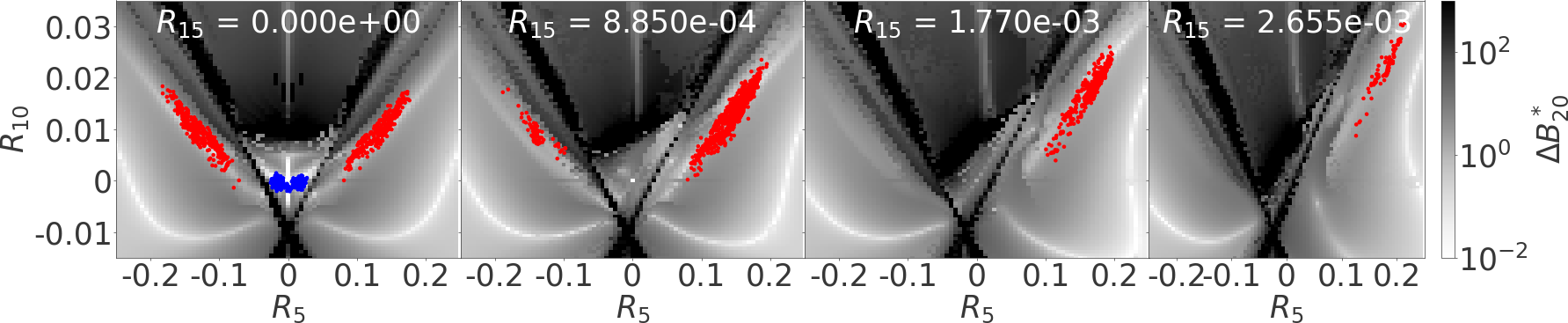}
    \caption{The top row shows the landscape of quasisymmetric stellarators for $n_{\text{fp}}=4$, the red dot on the leftmost panel corresponds to the HSX device. There are four quasisymmetry phases delineated by poorly quasisymmetric (dark color) devices. 
    Devices optimized for QA and QH in QUASR plotted on top of the landscape with blue and red dots, respectively.  The second row and third rows of plots correspond to $n_{\text{fp}}=4$ and 5, respectively. }\label{fig:landscape}
\end{figure}

To construct such a space, we start by providing the elements necessary to characterize a near-axis quasi-symmetric vacuum field \citep{Landreman_Sengupta_2019,rodriguez2023constructing}. Such a field is defined by providing the geometry of the magnetic axis (customarily given as a set of Fourier harmonics $\mathbf R, \mathbf Z$ describing the axis shape in cylindrical coordinates), and two scalars $\overline \eta$ and $B_{2c}$. Given such inputs, one can evaluate a function $\Delta B_{20}(\mathbf R, \mathbf Z, \overline \eta, B_{2c})$ which measures the magnitude of the symmetry breaking of $B$ at second order in the distance from the magnetic axis \citep{Landreman_Sengupta_2019,rodriguez2023constructing}. That is, if $\Delta B_{20}=0$, then the near-axis vacuum field considered has `perfect' quasisymmetry to second order. This scalar function of $(\mathbf R, \mathbf Z, \overline \eta, B_{2c}) \in \mathbb{R}^{2N+1}$ variables is too high-dimensional for straightforward plotting.
Using the approach in \cite{rodriguez2023constructing}, one may nevertheless reduce the dimensionality in such a way that the critical structure of the problem is retained. This reduction of the space draws from the following logic: 
\begin{enumerate}
    \item the main structural element in the field description is the axis shape, which will remain our free parameters. For visualization purposes, the axis geometry is restricted to a few harmonics (this is key only to visualization, but not to the concept itself), in this case $\mathbf R =  (1, R_{n_\text{fp}}, R_{2n_\text{fp}}, R_{3n_\text{fp}})$, and $\mathbf Z = (Z_{n_\text{fp}}, Z_{2n_\text{fp}}, Z_{3n_\text{fp}})$,
    \item for a given axis shape, $\overline \eta = \overline{\eta}^{\dag}(\mathbf R, \mathbf Z)$ is chosen uniquely to roughly minimize elongation of surfaces, formally
    \begin{equation}
        \etabar^\dag = \argmax_{\etabar} |\iota(\etabar) - N|.
    \end{equation}
    where $\iota$ is the on-axis rotational transform and $N$ is the helicity (see \cite{rodriguez2023constructing} for further details and proofs). Changing the value of $\Bar{\eta}$ will directly affect elongation of surfaces as well as the rotational transform.
    \item the $\mathbf Z$ harmonics and $B_{2c}$ are then optimized over to minimize $\Delta B_{20}$.
\end{enumerate}
This results in a scalar function
\begin{equation*}
    \Delta B^*_{20}(\mathbf R) = \min_{\mathbf Z, B_{2c}} \Delta B_{20}(\mathbf R, \mathbf Z, \overline \eta^{\dag}(\mathbf R, \mathbf Z), B_{2c})
\end{equation*}
of only a function of 3 independent variables $(R_{n_\text{fp}}, R_{2n_\text{fp}}, R_{3n_\text{fp}})$, and it can easily be visualized by evaluating it on a three-dimensional tensor product grid using the \texttt{pyQSC} code \citep{pyqsc}.  
The minimization problem over the $Z$ harmonics ($\mathbf Z$) uses local optimization algorithms, where the $Z$ harmonics provided to the optimization are initialized to have $\mathbf Z := \mathbf R$.  This results in magnetic axes with $Z$ harmonics that scale with $R$ as the optimization algorithm does not stray far from the initial guess.
The resulting plots have structure thanks to the topological properties of the magnetic axis \citep{rodriguez2022phases}.
This landscape can be divided into phases (or regions) where either QA and QH devices are found, Figure \ref{fig:landscape}, corresponding to axis shapes with different \textit{self-linking} numbers \citep{rodriguez2022phases,moffatt1992helicity,fuller1999geometric,oberti2016torus}.  
At the origin, we have the QA phase, surrounded by QH phases.
As the magnitude of the third harmonic increases (and thus the axis shaping increases), the QA region shrinks; there is a limit to how much an axis can be shaped before it gains a non-zero helicity and represent a QH device. The east and west QH regions have helicity $n_{\text{fp}}$, and the north and south regions have helicity $2n_{\text{fp}}$.

In \cite{rodriguez2023constructing}, two branches of QH stellarators with favorable quasisymmetry were observed: the HSX branch, as well as a new QH branch that had not yet been explored.
The HSX branch is the one on which the HSX device lies, and the `new QH' branch is the one below it (see the leftmost panel of the top row in Figure~\ref{fig:landscape}). To visualize the devices in QUASR on this landscape, we fit the near-axis model to the magnetic fields generated by the coil sets. The algorithm to do this is described in Appendix~\ref{sec:fitting}.

\subsection*{QUASR on the near-axis quasisymmetry landscape}

Following the procedure in Appendix~\ref{sec:fitting}, QUASR devices may be then approximated by near-axis fields (a form of reduced representation of the configuration) and represented in the reduced quasisymmetric phase-space discussed above. However, we must note that some of the simplifying assumptions behind such a reduced phase space plot in Figure~\ref{fig:landscape} do not necessarily hold for all the devices in QUASR. 
\par
First, note that the optimization that was used to design the devices in QUASR did not directly target the devices' elongation. As a result, we are not guaranteed that $\etabar^*$ will be close to $\etabar^\dag$ as the phase space construction assumes.  
Around 40\% of $n_{\text{fp}}=4,5$ devices have $|\etabar^*-\etabar^\dag|/|\etabar^\dag|<0.1$ (roughly a change of $\sim20\%$ in elongation) and this increases to around 70\% when the upper bound is relaxed to 0.2. This illustrates that there is a non-neglible portion of devices in QUASR with possibly unrealistic elongations, which motivates the addition of constraints on surface elongation to the design workflow in Figure \ref{fig:workflow}.
\par
Second, for illustration purposes, we focused on configurations whose axes had 3 dominant harmonics. A device is deemed to have $N$ dominant Fourier harmonics if $R_{k\,n_{\text{fp}}}(1+k^2n_{\text{fp}}^2), Z_{k\,n_{\text{fp}}}(1+k^2n_{\text{fp}}^2)<0.1$ for $k\geq N+1$. These conditions rely on rescaled Fourier harmonics that are dimensionless and naturally measure relevance of axis harmonics, introduced in \cite{rodriguez2022phases}. Around 50\% of devices have three dominant harmonics, and this number increases to around 70\% when four dominant harmonics are allowed. 
During the globalization of the near-axis phase of the workflow, the axis is described only using two Fourier harmonics. Then, the solution is subsequently polished and the number of harmonics used to describe the axis is increased.
This continuation procedure may have influenced these statistics.
Additionally, note that the choice to represent magnetic axes in cylindrical coordinates and Fourier harmonics might not be the optimal choice, and an alternative parametrization of the phase space could be better suited (even within the near-axis framework).

With the above in mind, we present on the phase space in Figure~\ref{fig:landscape} those devices in QUASR that satisfy: 
\begin{enumerate}
    \item  the device's $\etabar^*$ has a relative error of at most 10\% with respect to $\etabar^\dag$, i.e., $|\etabar^*-\etabar^\dag|/|\etabar^\dag| < 0.1$
    \item  the magnetic axis has only three dominant harmonics, i.e., $R_{k\,n_{\text{fp}}}(1+k^2n_{\text{fp}}^2), Z_{k\,n_{\text{fp}}}(1+k^2n_{\text{fp}}^2)<0.1$ for $k\geq4$.
\end{enumerate}

It is also ensured that all plotted devices have $R_{3n_{\text{fp}}}>0$.  If this is not the case, then the device is rotated by a half-period, which flips the sign of the odd harmonics: $R_{n_{\text{fp}}}, R_{3n_{\text{fp}}}$.
A device's coordinate on the landscape is computed from the harmonics of its magnetic axis.

We observe that the QA devices lie as expected in the QA region.  This QA region shrinks as the magnitude of the third harmonic increases, which corroborates the observation that QA devices typically do not have highly shaped magnetic axes. 
In the QH region, QUASR configurations appear to cluster around the HSX branch, noticeably leaving the `new QH' branch, which at the time of publication the authors in \cite{rodriguez2023constructing} noted that had not yet been explored. This is so even when optimization efforts were launched directly from the new branch, cases in which optimization trajectories left the branch to find better quasisymmetry elsewhere. Whether the struggle to find such configurations comes from the difficulty to find appropriate coils under the current engineering constraints, some intrinsic limitations on volumetric QS or the existence of a higher dimensional path in the phase space towards better QS, remains a subject of study.

While this comparison to the near-axis model is interesting and insightful, this approach has some important drawbacks.  
The landscape plots we obtain here are restricted to two or three dimensions before the visualizations become laborious. Despite this, it might still be possible to measure a device's distance to a favorable branch in higher dimensions, permitting some additional higher-dimensional insights. A better representation of the axes could also be fitting. 
In addition, the filters on $|\etabar^*-\etabar^\dag|/|\etabar^\dag|$ may also be unnecessarily restrictive, although as we have argued, this filter is not completely artificial if elongation of flux surfaces is taken into consideration.  
In the following section, we will use a different approach that applies to the entire data set.

\section{Principal component analysis}\label{sec:PCA}
Principal component analysis (PCA) is a dimensionality reduction technique that projects $D$-dimensional data points $\mathbf x_i \in \mathbb R^D$ onto a lower-dimensional manifold \citep{bishop2006pattern}. The vector $\mathbf x_i$ is also called a feature vector.  For the following explanation, we focus on reducing the dimensionality of the stellarators to two-dimensional points on the plane, though one is free to retain as many dimensions as desired.

The PCA finds a plane that lies closest to the data points, and then projects the data points on this plane, thereby reducing its dimensionality.
Any projected point can be expressed as $\alpha_1 \mathbf d_1 + \alpha_2 \mathbf{d}_2 + \bm \gamma$, for some $\alpha_1, \alpha_2 \in \mathbb R$, where $\mathbf d_1$ and $\mathbf d_2$ are orthonormal.
The vectors $\mathbf d_1, \mathbf d_2$ and $\bm \gamma$ are chosen to minimize the error:
\begin{equation}\label{eq:min}
   \bm \gamma^*, \mathbf d^*_1, \mathbf d^*_2 = \argmin_{\substack{ \bm \gamma^, \mathbf d_1, \mathbf d_2\\
   \text{s.t.} \| \mathbf d_1\|, \| \mathbf d_2\| = 1}}  \frac{1}{N} \sum_{i=1}^N \| \mathbf x_i  - \tilde{\mathbf x}_i \|^2,
\end{equation}
where $\tilde{\mathbf x}_i$ is the projection of the original point $\mathbf x_i$ onto the plane of the form
\begin{equation*}
    \tilde{\mathbf x}_i = \bm \gamma + \alpha_{1,i} \mathbf d_1 + \alpha_{2,i} \mathbf d_2.
\end{equation*}
From the orthonormality of $\mathbf d_1$ and $\mathbf d_2$, it can be shown that $\alpha_{1,i} = \mathbf d_1 \cdot ( \mathbf x_i-\bm \gamma)$ and $\alpha_{2,i} = \mathbf d_1 \cdot ( \mathbf x_i-\bm \gamma)$, and the optimal choice of $\bm\gamma$ is the mean of the data i.e. $\bm \gamma^* =\overline{\mathbf x}= \frac{1}{N}\sum_{i = 1}^N\mathbf x_i$.  The Karush-Kuhn-Tucker conditions of the minimization problem \citep{bishop2006pattern} in \eqref{eq:min} reveal that the optimal choice of vectors $\mathbf d^*_1, \mathbf d^*_2$ are given by the first two eigenvectors of the covariance matrix 
\begin{equation*}
 \frac{1}{N}   \sum^N_{i=1} (\mathbf x_i - \overline{\mathbf x})^T (\mathbf x_i - \overline{\mathbf x}).
\end{equation*}
The vectors $\mathbf d^*_1, \mathbf d^*_2$ are called respectively the first and second principal components.
For this choice of $\bm \gamma^*, \mathbf d^*_1, \mathbf d^*_2$, the reconstruction error is 
\begin{equation*}
    \frac{1}{N} \sum_{i=1}^N \| \mathbf x_i  - \tilde{\mathbf x}_i \|^2 = \lambda_3 + \cdots + \lambda_D,
\end{equation*}
which is the sum of the $D-2$ remaining eigenvalues of the covariance matrix.  The ratios $\lambda_1/(\lambda_1 + \cdots + \lambda_D)$ and $\lambda_2/(\lambda_1 + \cdots + \lambda_D)$ measure by how much the first two principal components reduced the reconstruction error.  If the sum of these two quantities is close to 1, then little information was lost in the dimensionality reduction.  One is free to include as many principal components as desired, but for simplicity of plotting and ease of understanding, we restrict ourselves to two or three principal components.

\begin{figure}
    \centering
    \begin{tikzpicture}
        \node (a) at (0,0) {\includegraphics[width=0.6\linewidth]{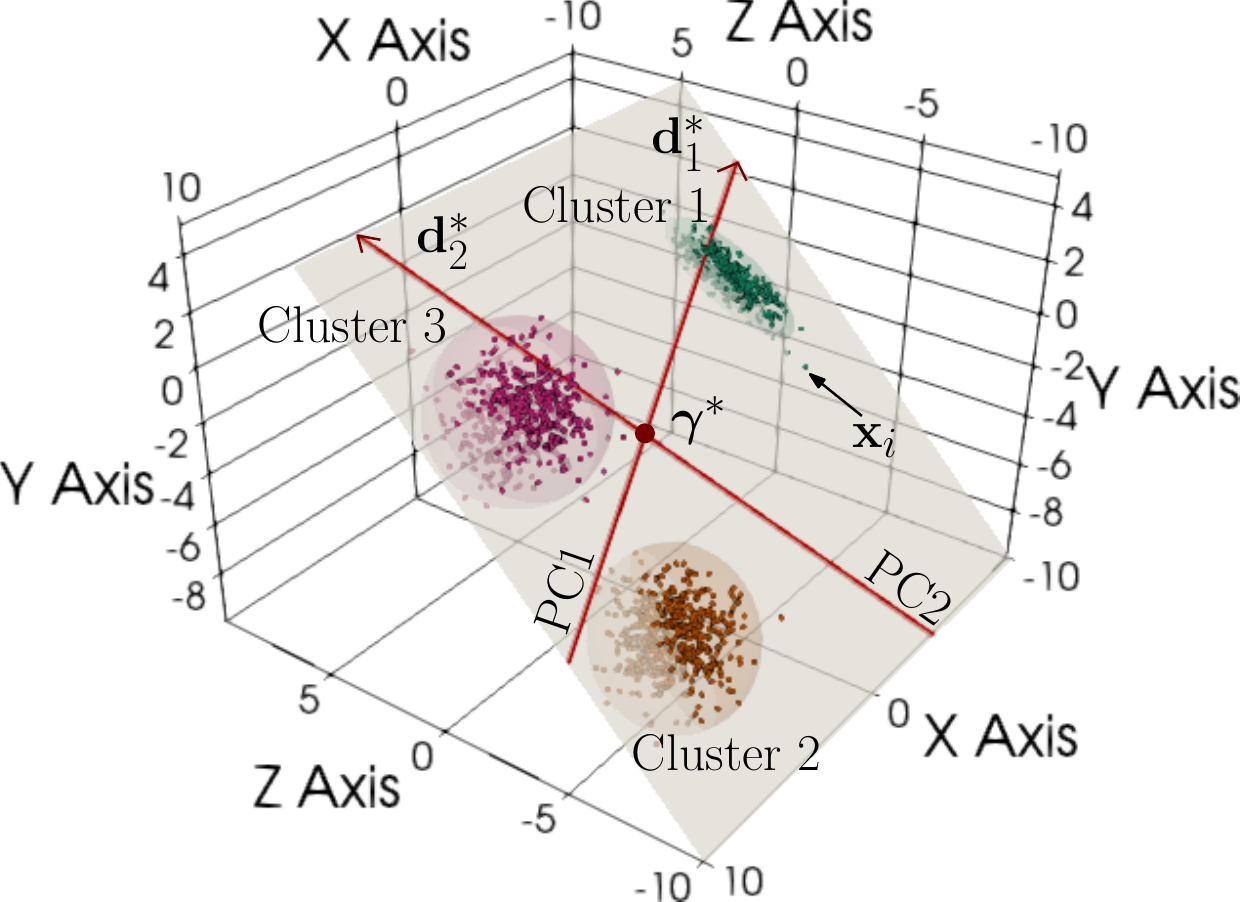}};
        \node (b) at (6.75,0) {\includegraphics[width=0.35\linewidth]{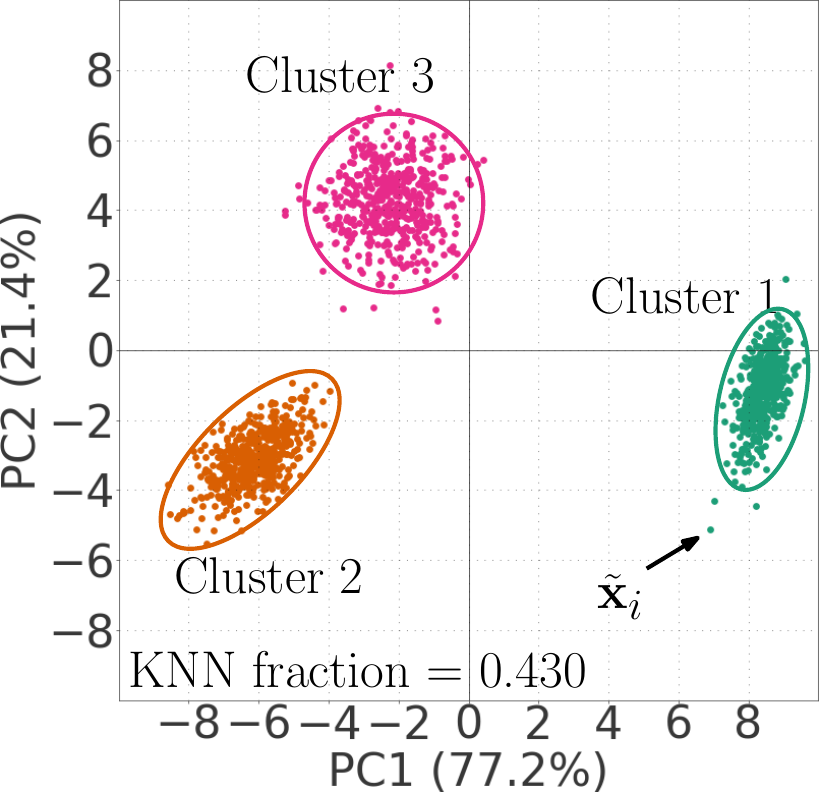}};
    \end{tikzpicture}
    
    \caption{Data $\mathbf x_i \in \mathbb R^3$ from three Gaussians are shown on the left, and the optimal projection $\tilde{\mathbf x}_i$ onto a two-dimensional plane is shown on the right.
    The transparent ellipsoids correspond to the 95\% confidence interval of each Gaussian.
    }
    \label{fig:gaussians}
\end{figure}

Once a PCA is completed, each high-dimensional data point $\mathbf x_i$ will have a corresponding two-dimensional coordinate $(\mathbf d^*_1 \cdot ( \mathbf x_i-\bm \gamma^*), ~\mathbf d^*_2 \cdot ( \mathbf x_i-\bm \gamma^*)) \in \mathbb R^2$.
We illustrate this procedure with a three-dimensional set of data points $\mathbf x_{i} \in \mathbb R^3$ on the left of Figure \ref{fig:gaussians}, projected down to a two-dimensional plane on the right.  
Projected points far away from each other on the plane will also be far away from each other in the higher dimensional space. In other words, dissimilar devices in the low-dimensional representation are dissimilar in higher dimensions.  Note that the converse of this statement is not necessarily true.

The advantages of PCA is that it is an unsupervised technique that does not require labels or prior knowledge about the data. It identifies important features in a low-dimensional subspace that describe the data set and permit its visualization. By doing so, it reduces the complexity and noise in the data and gives an idea of its intrinsic dimensionality, which is revealed by examining the eigenvalues of the covariance matrix. PCA is often used as a pre-processing step for machine learning and optimization algorithms to expose the most important components of the data \citep{kobak2019art}.  It also does not have any parameters to adjust, which makes it straightforward to apply to a data set.

However, PCA can also have disadvantages. By reducing the dimensionality of the data, we may lose some important information.
The transformation to low-dimensional space may distort the geometric relationships that existed in the original space. 
In addition, PCA is useful when the data is well approximated on a linear manifold of dimension less than $D$, which is certainly not always the case.  For example, consider a data set that lies on a three-dimensional helix.  The intrinsic dimension of the data is one, since the helix can be unraveled to parameterize any point on it by a single coordinate.  But applying PCA to the data set distorts it, since its topology is ill-approximated on a linear manifold of dimension less than three.
In cases like this, nonlinear dimensionality reduction techniques can be useful, and the literature on these approaches is rich.  For example, isomap \citep{tenenbaum2000global}, and t-Distributed Stochastic Neighbor Embedding \citep{van2008visualizing} are all well-known algorithms to accomplish this task.
The disadvantage of nonlinear techniques is that they may have user-defined tuning parameters, which can be difficult to select.
Since the underlying topology of our data set is not known, and to avoid the difficulties associated with nonlinear techniques, we use PCA in this work.
 Auto encoders have also been proposed as interesting methods for compression or dimensionality reduction. \citep{bourlard1988auto} showed that auto encoders perform a similar operation to SVD or low rank matrix approximation. Since PCA is a type of low rank matrix approximation, this is an interesting connection to what is done in this work.

In light of the above discussion, we use a global and local measure of accuracy to evaluate how faithful the data in lower-dimensions is to the original:
\begin{enumerate}
    \item \textit{the cumulative PC ratios}: as mentioned above, this score quantifies how much information was lost in the projection onto the lower-dimensional linear manifold, corresponding to the sum of the PC ratios:  $\lambda_1/ (\lambda_1 + \hdots + \lambda_{D})$ when projecting onto a 1D manifold,  $(\lambda_1 + \lambda_2 )/ (\lambda_1 + \hdots + \lambda_{D})$ when projecting onto 2D manifold, and $(\lambda_1 + \lambda_2 + \lambda_3)/ (\lambda_1 + \hdots + \lambda_{D})$ when projecting onto a 3D manifold.  The closer this value is to 1, the less information is lost in the projection.  If more PCs are included, then the corresponding eigenvalues are added to the numerator.  This is a global measure of accuracy.
    \item \textit{KNN fraction}: the fraction of $k$-nearest neighbors in the high-dimensional data set that remain in the two-dimensional projected data points \citep{lee2009quality, kobak2019art}, called the `KNN fraction'.  
    The purpose of this quantity is to measure how accurately local relationships are preserved after dimensionality reduction. 
    Note that for very large $k$, this fraction approaches 1, so some reasonable choice of $k$ has to be picked.
    This is a local measure of accuracy.
\end{enumerate}
We present these at work in a simple example of Gaussian processes in Appendix~\ref{app:gaussians}. In what follows, we define how we represent the different devices in the QUASR dataset, and then analyze them using the tools above. 

\subsection{Feature vectors}
In order to apply PCA to our set of stellarators, we must define how a device's feature vector is constructed.
A device's feature vector is defined to have $D=663$ entries, where each entry is a Fourier harmonic of a stellarator symmetric surface in the device's vacuum magnetic field, using $m_{\text{pol}}, n_{\text{tor}}=10$, where $m_{\text{pol}}, n_{\text{tor}}$ are respectively the number of poloidal and toroidal harmonics used to represent the surface geometry. We use the special surface representation \citep{surfaceopt1}:
\begin{align}
         x(\varphi, \theta) &= \cos(\varphi)\hat x(\varphi, \theta) -  \sin(\varphi)\hat y(\varphi, \theta), \label{eq:p1}\\
         y(\varphi, \theta) &= \sin(\varphi)\hat x(\varphi, \theta) +  \cos(\varphi)\hat y(\varphi, \theta),  \\ 
         z(\varphi, \theta) &= \sum_i \sum_j z_{i, j} u_i(\varphi) v_j(\theta),
\end{align}
where
\begin{align}
    \hat x(\varphi, \theta) &= \sum_i \sum_j \hat x_{i, j} u_i(\varphi) v_j(\theta), \\
    \hat y(\varphi, \theta) &= \sum_i \sum_j \hat y_{i, j} u_i(\varphi) v_j(\theta), \label{eq:p5}
\end{align}
and
\begin{align*}
    u(\varphi) &= (1, \cos(2\pi n_{\text{fp}} \varphi), \sin(2\pi n_{\text{fp}}\varphi), \hdots, \cos(2\pi n_{\text{fp}} n_{\text{tor}}\varphi), \sin(2\pi n_{\text{fp}}n_{\text{tor}}\varphi))\\
    v(\theta) &= (1, \cos(2\pi  \theta), \sin(2\pi \theta), \hdots, \cos(2\pi m_{\text{pol}}\theta), \sin(2\pi m_{\text{pol}}\theta)).
\end{align*}
Since we are only working with stellarator symmetric devices, the symmetry-breaking basis functions in the sums of \eqref{eq:p1}-\eqref{eq:p5} are skipped.
This representation uses three truncated Fourier series to represent the Cartesian coordinates of the surface: $(x(\varphi, \theta), y(\varphi, \theta), z(\varphi, \theta))$.
This is in contrast to VMEC \citep{hirshman1983} and DESC \citep{Panici_Conlin_Dudt_Unalmis_Kolemen_2023}, which use a cylindrical representation $(R(\phi, \theta), \phi, Z(\phi, \theta))$, with $\phi$ the standard cylindrical toroidal angle.

\begin{figure}
    \centering
\includegraphics[height=0.15\textheight, trim=0 0 50 0, clip]{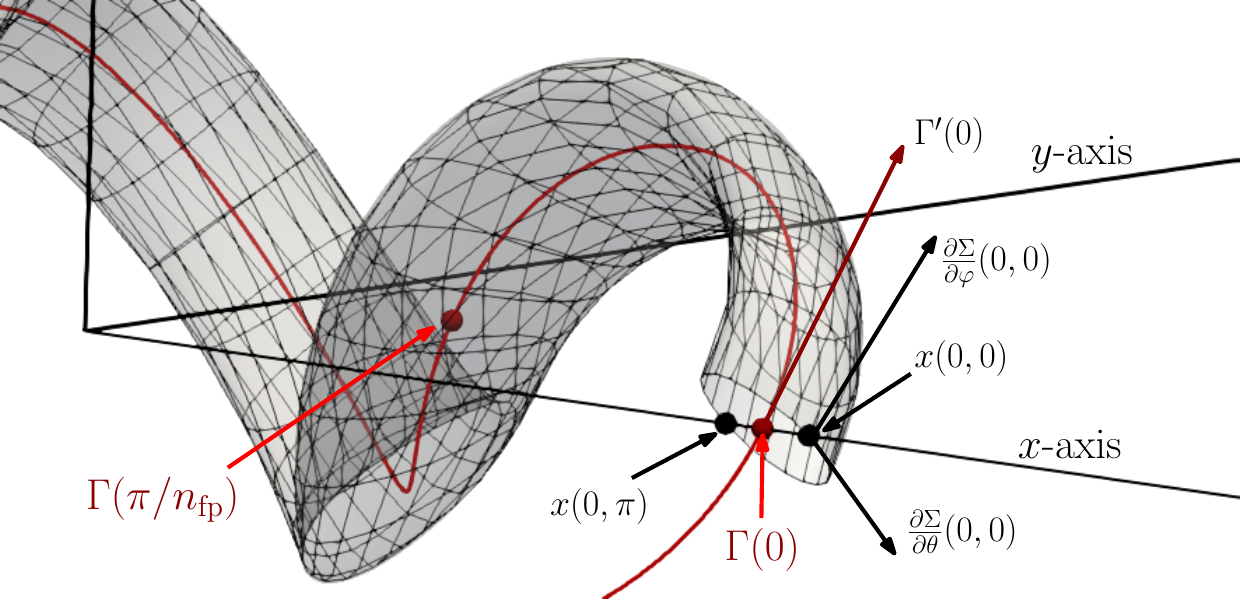}
\includegraphics[height=0.15\textheight, trim=0 0 100 0, clip]{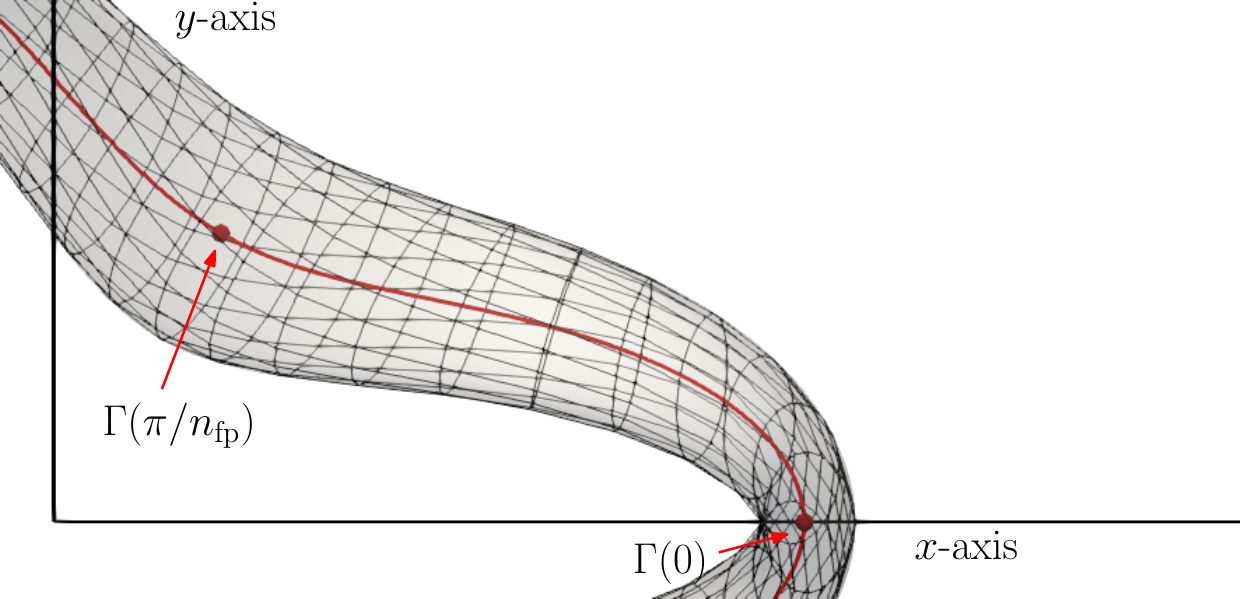}
    \caption{Conditions that the surface parametrization must satisfy so that the feature vector of Fourier coefficients is unique.  The left image illustrates how the surface coordinates satisfy $x(0, 0) \geq x(0, \pi)$, $\frac{\partial z}{\partial \theta}(0, 0) \geq 0$, $\frac{\partial y}{\partial \varphi}(0, 0) \geq 0$, and that the $Z$-coordinate of the magnetic axis (in red) satisfies $Z'(\phi) \geq 0$.  The right image is a view in the $\sm Z$ direction onto the $XY$ plane, showing that the radius of the magnetic axis satisfies $R(0) \geq R(\pi/n_{\text{fp}})$.
    The grid lines on the magnetic surface correspond to lines of constant toroidal and poloidal Boozer angles.}
    \label{fig:fv}
\end{figure}

The feature vector associated to a device contains all the Fourier harmonics $\hat x_{i, j}, \hat y_{i, j}$ and $\hat{z}_{i, j}$ associated to the surface, without any additional normalization. We emphasize this lack of normalization because in the context of PCA it is usually recommended to normalize the data across samples. One common normalization is to rescale each component of the feature vector to a Gaussian with mean zero and covariance one to ensure no feature dominates the analysis due to it being much larger. However, this kind of normalization only makes sense if the different components of the feature vector are not meaningfully related to each other and their values have different orders of magnitude because they come from inherently different quantities or phenomena. 
In our case, the feature vectors contain Fourier coefficients, which decay as the harmonic increases. Therefore, we chose to forgo the normalization of different Fourier components across the devices because the values of the Fourier coefficients have a very important meaning: they are related to each other to describe a single surface. 
For the vacuum field devices in QUASR, the magnetic field of the device is fully specified by the geometry of a single magnetic surface, regardless of how this is parametrised.   Additional parameters would be required in the feature vector for devices with plasma pressure, and normalization might be indicated there.
This leaves some non-uniqueness in our chosen representation, as different angle choices for the defining surface can lead to different fourier coefficients defining the same magnetic surface geometry. To remove this indeterminacy, we choose the angles on the surface $(\varphi, \theta)$ to correspond to Boozer coordinates. Even then, there remains certain non-uniqueness; for example, the same configuration rotated by a half-period about the $Z$-axis, or reflected about the $XY$ plane would correspond to different feture vectors. Thus, to ensure that geometrically similar devices are close to each other in Fourier space, in addition to 
\begin{enumerate}
    \item[1.] making the toroidal and poloidal angles on the surface Boozer angles $(\varphi, \theta)$,
\end{enumerate}
we check that the magnetic axis in cylindrical coordinates $\Gamma(\phi) = (R(\phi), \phi, Z(\phi))$ satisfies:
\begin{enumerate}
    \item[2.] $Z'(0) \geq 0$,
    \item[3.] $R(0) \geq R(\pi/n_{\text{fp}})$,
\end{enumerate}
and that the surface $\Sigma(\varphi, \theta) = (x(\varphi, \theta), y(\varphi, \theta), z(\varphi, \theta))$ in Boozer coordinates wraps toriodally and poloidally such that:
\begin{enumerate}
    \item[4.] $x(0, 0) \geq x(0, \pi)$,
    \item[5.] $\frac{\partial z}{\partial \theta}(0, 0) \leq 0$,
    \item[6.] $\frac{\partial y}{\partial \varphi}(0, 0) \geq 0$.
\end{enumerate}
These conditions are illustrated in Figure \ref{fig:fv}.
The second and third conditions break the freedom about the stellarator symmetric points.  The final three conditions ensure that the origin of the surface $(\varphi=0, \theta=0)$ lies on the outboard side of the stellarator, that the parametrization wraps poloidally clockwise if looking in the $+Y$ direction at the curve given by $\Sigma(\theta, 0)$, and that increasing $\varphi$ means going counterclockwise about the $Z$-axis if looking in the $-Z$ direction.
If these conditions are not satisfied, we reflect the device about the $XY$ plane, rotate it by a half-period about the $Z$-axis, or reparametrize the surface so that it wraps in the correct direction. 
The direction of the inequalities in conditions 2-6 were arbitrarily chosen.

A more compact feature vector could have been generated by converting the representation in \eqref{eq:p1}-\eqref{eq:p5} to a cylindrical one, thereby reducing the number of Fourier series from three to two.  
We do not do so here because the coil design workflow relies on the Cartesian representation of the surfaces. Although Cartesian coordinates are used in the coil design workflow, this does not mean any field shape is possible. Note that in Phase I (the near-axis phase of the coil design) it is assumed that the magnetic axis can be represented in cylindrical coordinates. As a result, we will not find devices with more general magnetic axis geometries, e.g., devices where the axis is anywhere vertical, or knotted. Generalizing our implementation is the subject of future work.

In addition, this conversion is not lossless when restricted to a finite dimensional Fourier series, as explained in section \ref{sec:qh_nfp4_iota2p3}. Other forms of parametrization could also be possible. It is tempting, for example, to represent each configuration by its second order near-axis model, which constitutes a significantly reduced space. Although this could be interesting, it does not correspond to a one-to-one map, and thus should not be used as a unique representation. This is because, a priori, one could devise an infinity of distinct fields, all of which correspond to the same second order near-axis construction. Another alternative could be to use a vector of Fourier harmonics associated to the coils and their currents.  We did not do this due to the inherent non-uniqueness of coil sets, i.e., multiple, vastly different coil sets can produce similar magnetic fields. That being said, we have not fully explored these directions, and it might be interesting to do this in the future.

\begin{figure}
    \centering
    \includegraphics[width=\textwidth, trim=0 0 20 45, clip]{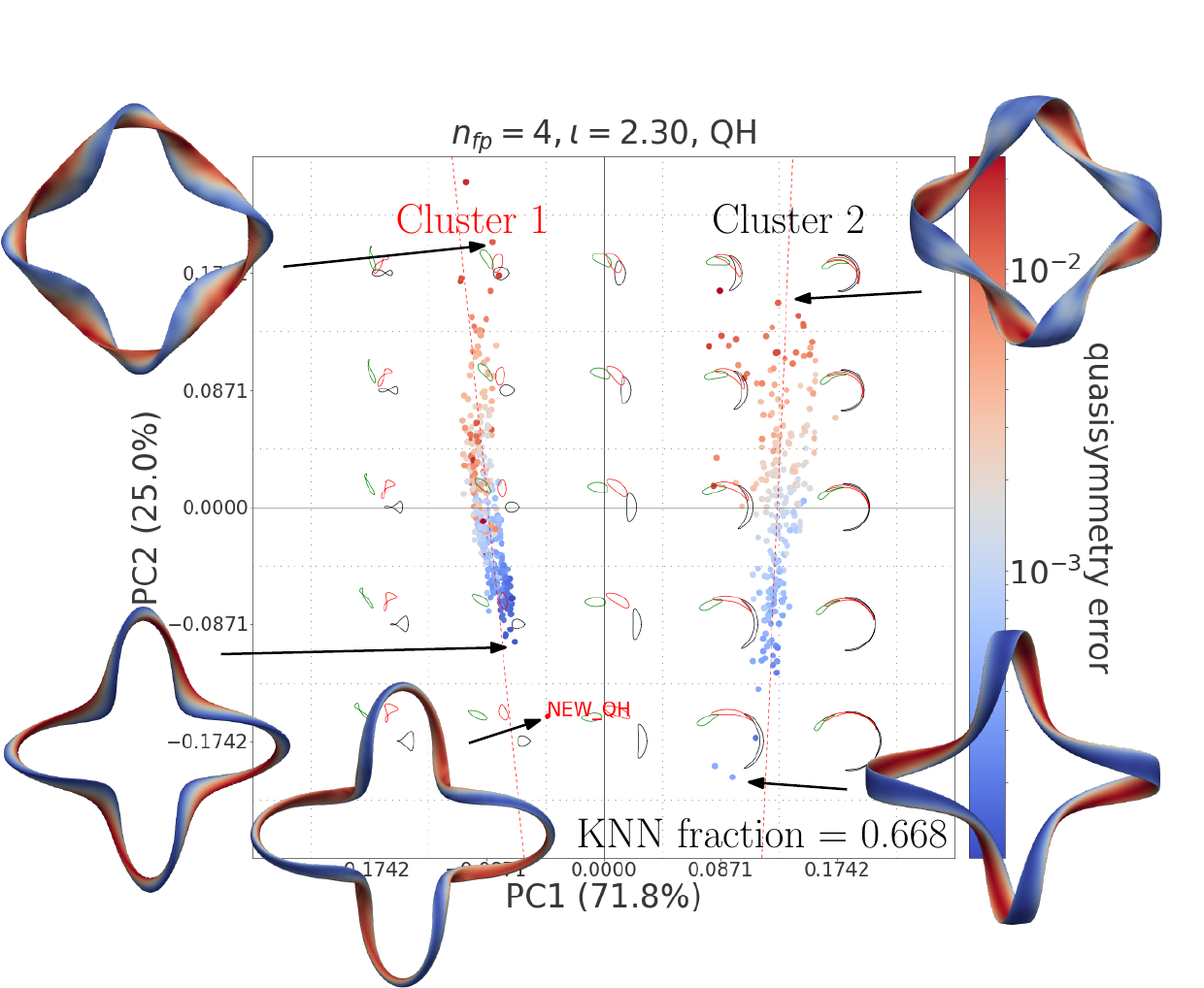}
    \caption{PCA of quasihelically symmetric devices in QUASR, along with the new QH device indicated with red text. At the center of each dashed square, we take the associated surface DOFs and generate cross sections to illustrate the geometric diversity of devices.  This type of cross section diagram is common in morphometrics literature \citep{bonhomme2014momocs}.  The red dashed lines correspond to projections of one-dimensional PCAs of the left and right clusters onto the two-dimensional manifold.
    The color on the magnetic surfaces corresponds to the local field strength.
    The quality of quasisymmetry is calculated by taking the square root of the average of \eqref{eq:qa1} on a number of surfaces of the magnetic field.
    }
    \label{fig:pca_newQH}
\end{figure}

\par
When presenting QUASR devices in what follows, we will do so alongside quasisymmetric devices available in the literature (see details in Appendix~\ref{app:configs}).

\subsection{Quasihelical symmetry ($n_{\text{fp}}=4, \iota=2.3$)} \label{sec:qh_nfp4_iota2p3}
To start, we look at $n_{\text{fp}}=4$ devices with a mean rotational transform of 2.3 in Figure \ref{fig:pca_newQH}.  
We observe that there are two distinct clusters of devices.  Within these clusters, the devices form a continuum with varying qualities of quasisymmetry.  

One may question how faithful this two-dimensional plot is to the true high-dimensional data set. 
A hint that this plot is an accurate representation of the relationships in the data is the sum of the percentages in the $x$ and $y$ axis labels is large.  This is evidence that including a third dimension is likely unnecessary to understand how devices relate to each other, which can be confirmed via a three-dimensional PCA.
Using 10 nearest neighbors in the KNN calculation, the average KNN fraction over all data points is 0.689.  This indicates that on average, just under 70\% of each node's 10 nearest neighbors in the high-dimensional representation are preserved on the lower dimensional manifold.
These projection error measures are comparable to the dimensionality reduction example of data sampled from simple Gaussians, in the introduction of Appendix~\ref{app:gaussians}.

Each point on the landscape plot corresponds to a specific vacuum magnetic field, with associated magnetic surfaces.  There are box-shaped regions delineated by dashed lines on the plot.  Taking the point at the center of each box, we compute cross sections of the associated magnetic surface.  The cross sections are plotted in black, red, and green, associated to the cylindrical angle $\phi=0, \pi/(4n_{\text{fp}}), \pi/(2n_{\text{fp}})$, respectively.
This figure makes it clear how devices differ from each other on the plane, and importantly how moving along the principal components affects the geometry of the device. Moving in the positive direction associated to $\mathbf d^*_1$, the elongation appears to increase, while going in the negative direction the elongation decreases.  In the positive $\mathbf d^*_2$ direction, the relative positions of the cross sections move, and in particular the axis excursion increases.
These cross sections do not necessarily correspond to physically relevant devices, as illustrated by the leftmost column, where the cross sections are self-intersecting.  This is not surprising, as the devices lie on a nonlinear manifold, and the PCA slices through it with a linear one.  If one strays too far away from the origin of the plot (the sample mean), the approximation errors become noticeable and unphysical devices appear.

\begin{figure}
    \centering
    \includegraphics[width=\textwidth]{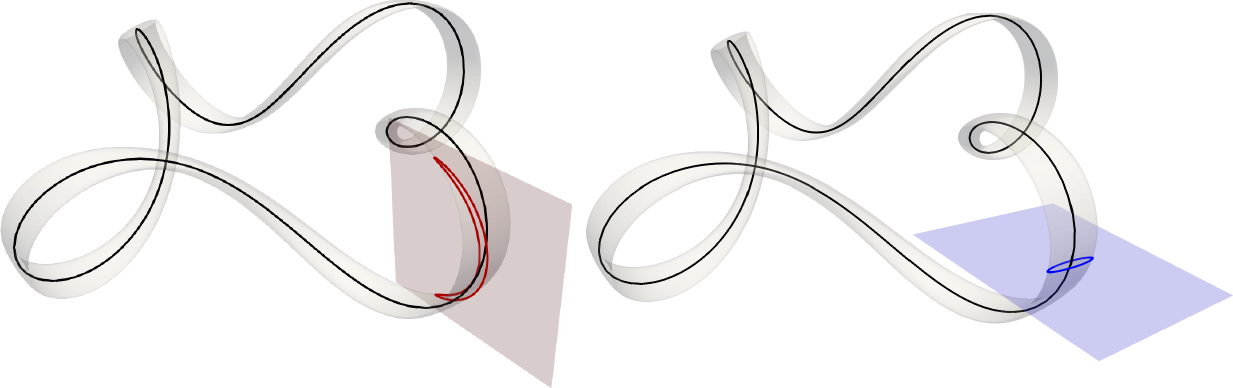}
    \caption{Cross sections of a device from cluster 2 in Figure \ref{fig:pca_newQH}.  The section is computed in the $RZ$ plane (red) and the plane orthogonal the magnetic axis (blue).  The magnetic axis is drawn in black.  Both planes pass through the same point on the magnetic axis.}
    \label{fig:xs}
\end{figure}

Within the space populated by the QUASR devices, we find that converting the highly-shaped surfaces, e.g. the device in Figure \ref{fig:xs}, to a cylindrical representation, as is used in the more standard equilibrium solvers VMEC or DESC, requires more than $m_{\text{pol}},n_{\text{tor}}=20$ Fourier modes. This is significantly larger than the number of modes used to design the stellarator ($m_{\text{pol}},n_{\text{tor}}=10$) with the representation in equation \eqref{eq:p1}-\eqref{eq:p5}.
Figure~\ref{fig:xs} illustrates the geometrical origin of this difficulty, where due to the large helical excursion of the magnetic axis, cross sections in the $RZ$ plane become highly elongated, leading to a misleading interpretation of the elongation of the device. This occurrence is particularly true for the devices on the right branch, in the bottom quadrant, where standard equilibrium solvers such as VMEC experience difficulties. These problems are probably shared by DESC \citep{Panici_Conlin_Dudt_Unalmis_Kolemen_2023} which uses the same cylindrical representation, but could perhaps be alleviated in the Frenet version of the GVEC code \citep{hindenlang2024generalized}. 

As a reference for this set, we also provide the location of the \texttt{NEW\_QH} device from section \ref{sec:nae_landscape} and observe again that there are no devices in this subset of QUASR that are close to it.
The \texttt{NEW\_QH} device appears closest to the low elongation branch, in the vicinity of devices with excellent quasisymmetry, but is not contained in QUASR.

\subsubsection*{Visualizing the continuum of devices in each cluster}
\begin{figure}
    \centering
    \resizebox{\textwidth}{!}{
    \begin{tikzpicture}
        \node (A) at (0,0) {\includegraphics[width=0.25\textwidth, trim=40 40 40 40, clip]{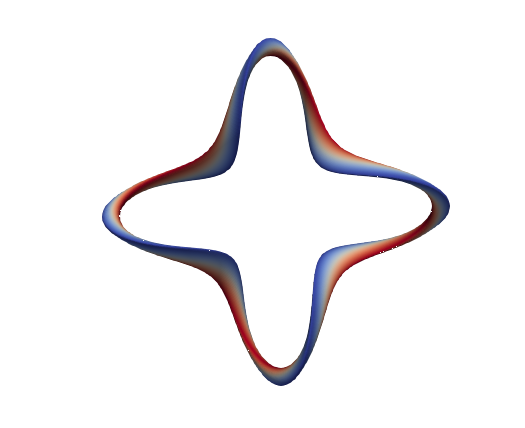}};
        \node (B) at (3.2,0) {\includegraphics[width=0.25\textwidth, trim=40 40 40 40, clip]{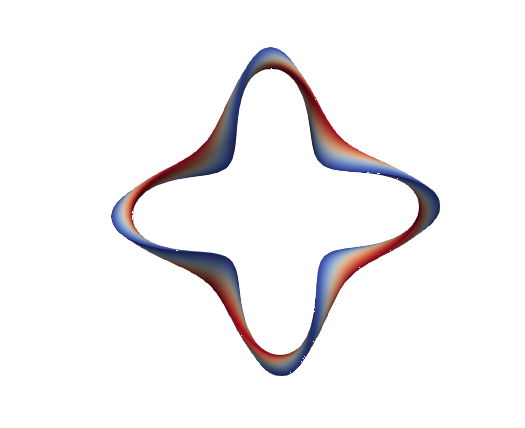}};
        \node (C) at (6.2,0) {\includegraphics[width=0.25\textwidth, trim=40 40 40 40, clip]{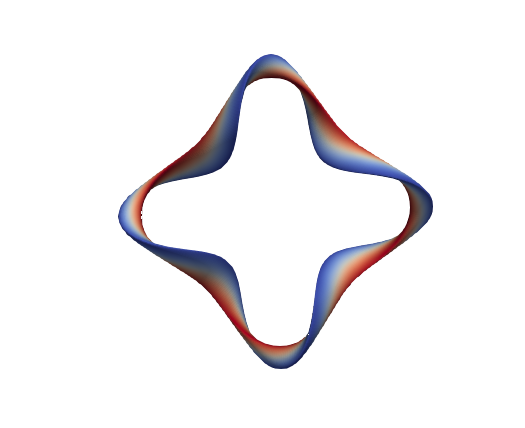}};
        \node (D) at (9.2,0) {\includegraphics[width=0.25\textwidth, trim=40 40 40 40, clip]{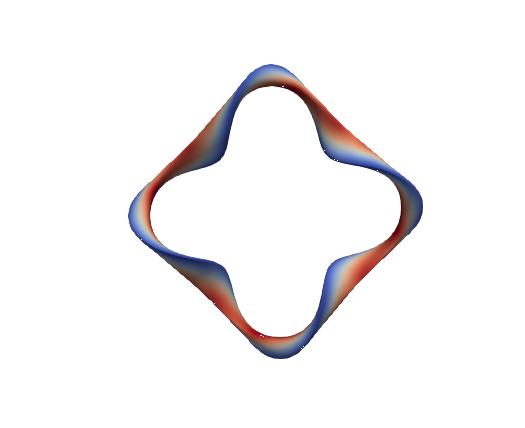}};
        \node (E) at (12.2,0) {\includegraphics[width=0.25\textwidth, trim=40 40 40 40, clip]{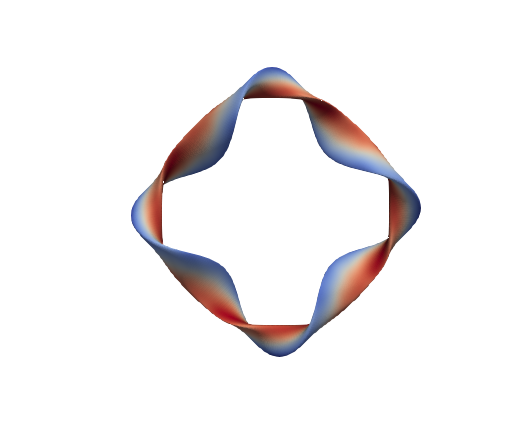}};

        \node (A) at (-1.2, -2) {PC = };
        \node (A) at (0, -2) {$-0.101$};
        \node (A) at (3, -2) {$-0.014$};
        \node (A) at (6.2, -2) {$0.073$};
        \node (A) at (9.2, -2) {$0.160$};
        \node (A) at (12.2, -2) {$0.248$};
        \node (A) at (-1.55, -2.5) {QS error = };
        \node (A) at (0, -2.5) {1.332e-04};
        \node (A) at (3, -2.5) {9.232e-04};
        \node (A) at (6.2, -2.5) {2.338e-03};
        \node (A) at (9.2, -2.5) {9.930e-03};
        \node (A) at (12.2, -2.5) {1.903e-02};

        \node (A) at (0,-4.5) {\includegraphics[width=0.22\textwidth, trim=50 50 50 50, clip]{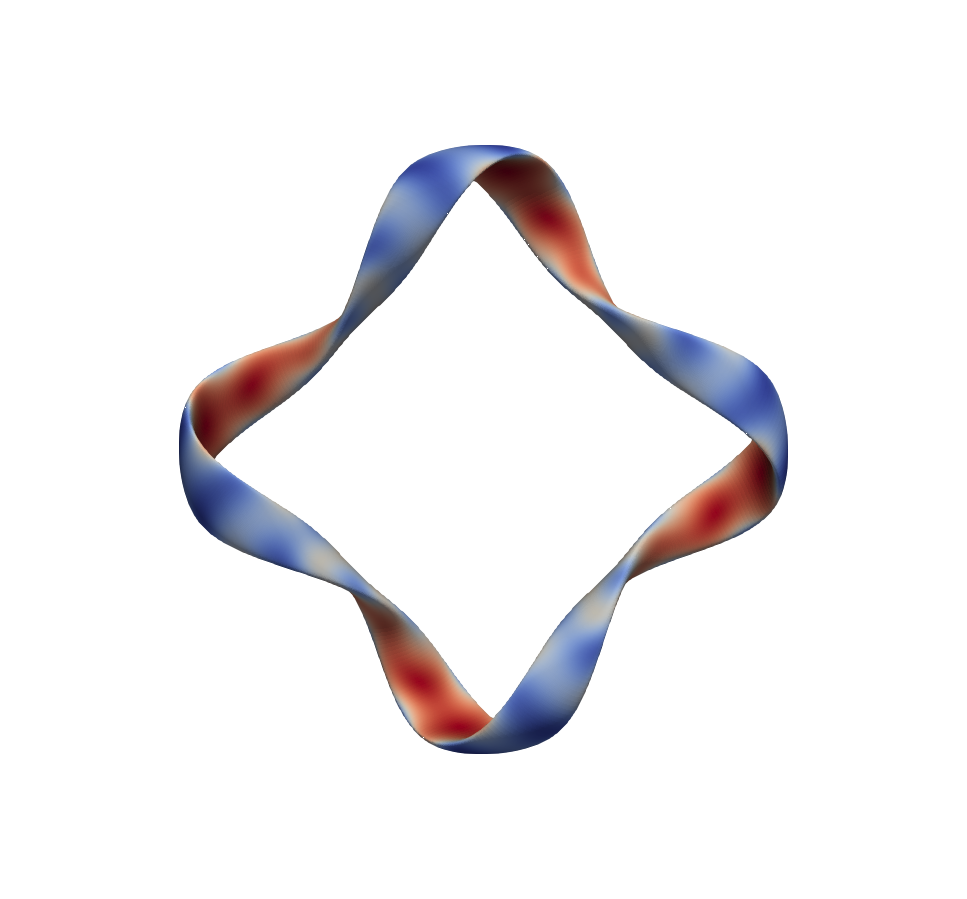}};
        \node (B) at (3.2,-4.5) {\includegraphics[width=0.22\textwidth, trim=50 50 50 50, clip]{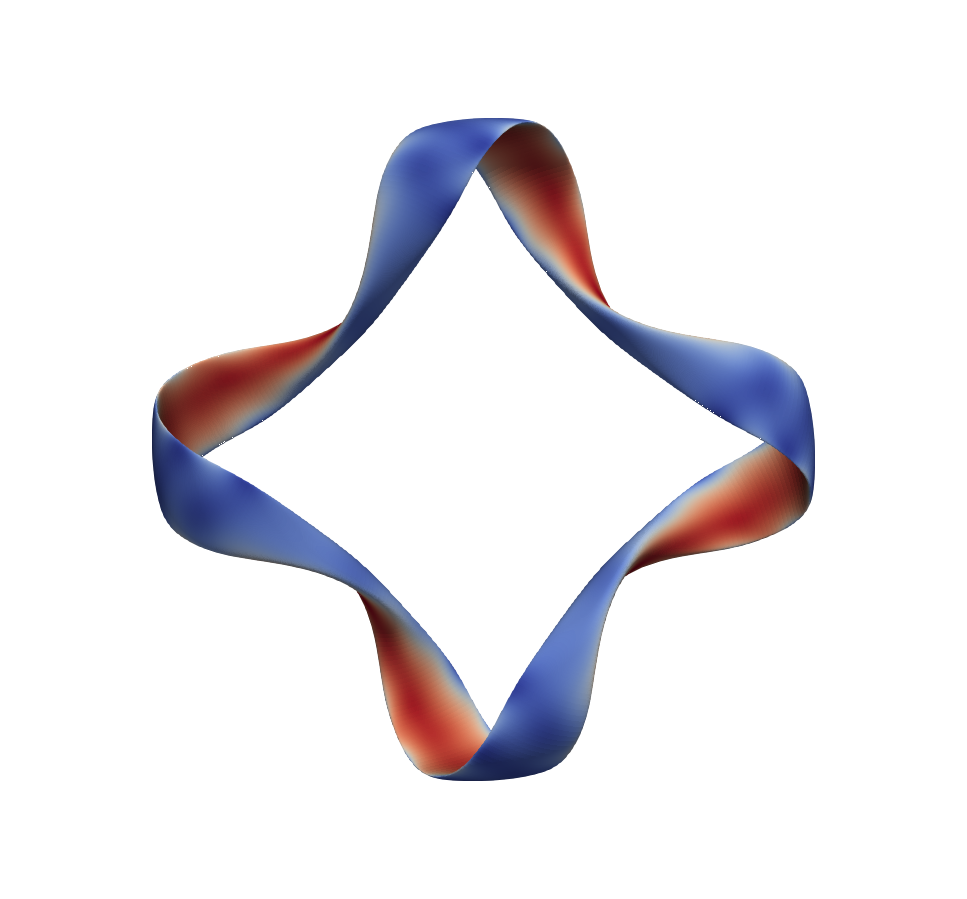}};
        \node (C) at (6.2,-4.5) {\includegraphics[width=0.22\textwidth, trim=50 50 50 50, clip]{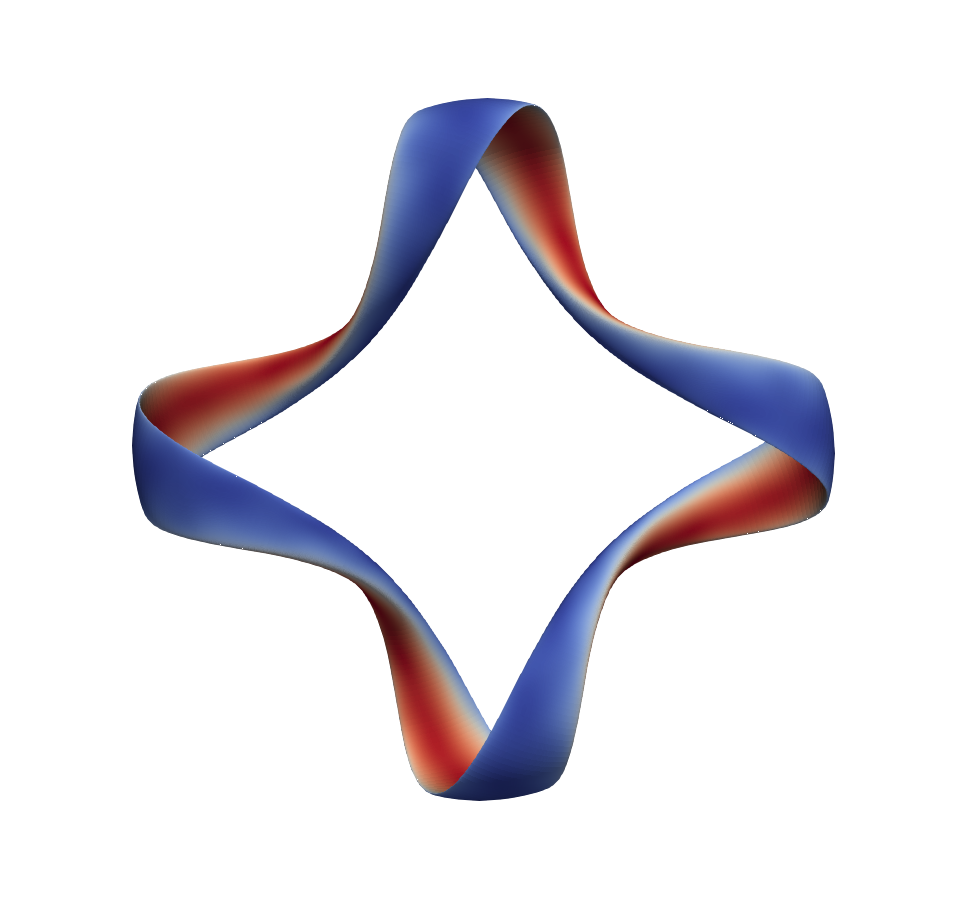}};
        \node (D) at (9.2,-4.5) {\includegraphics[width=0.22\textwidth, trim=50 50 50 50, clip]{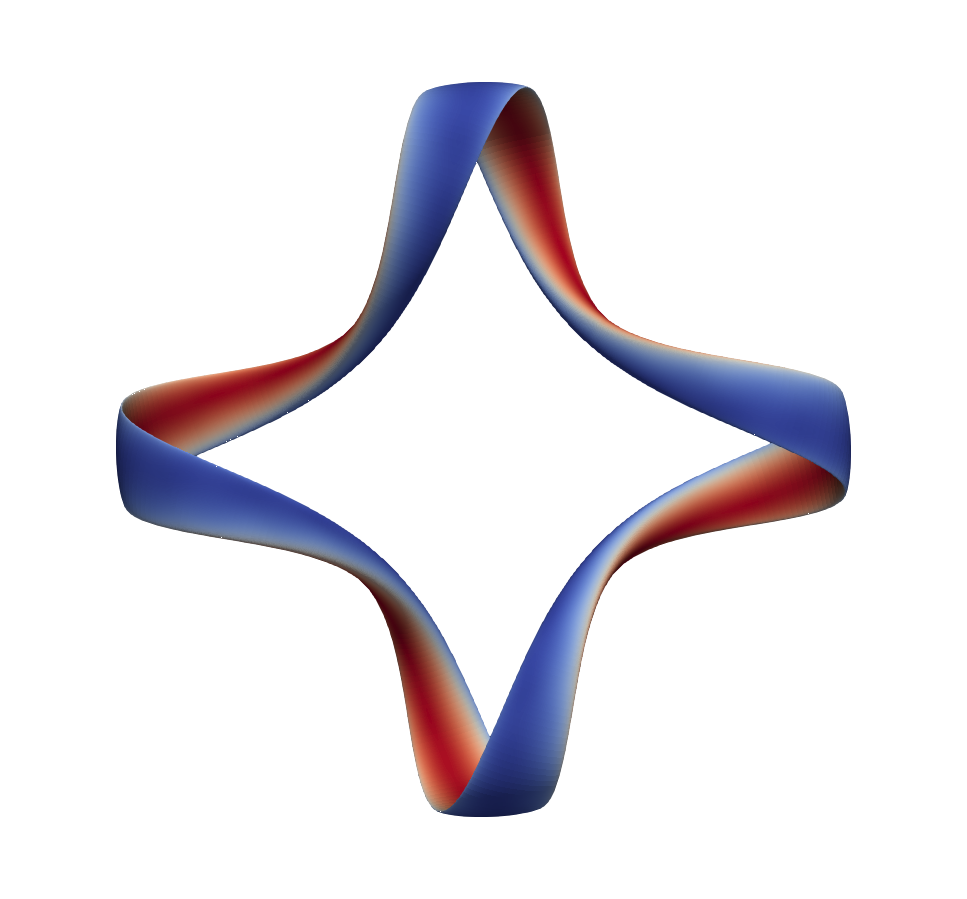}};
        \node (E) at (12.2,-4.5) {\includegraphics[width=0.22\textwidth, trim=50 50 50 50, clip]{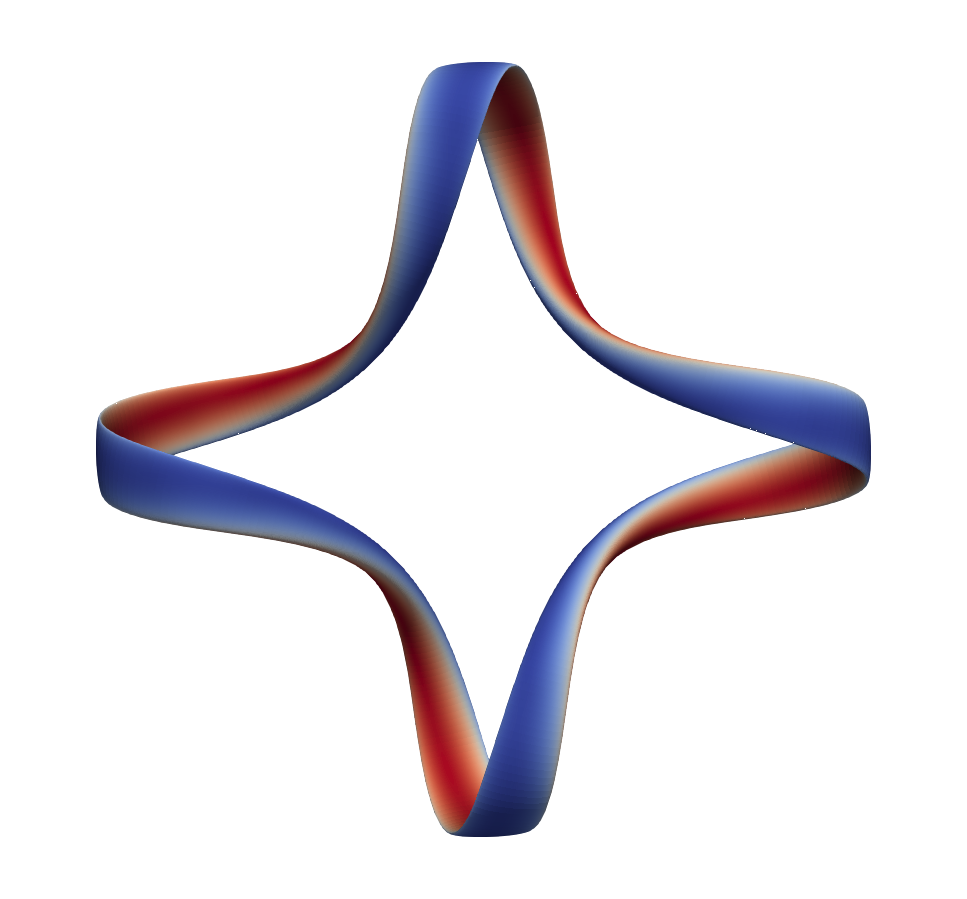}};
        
        \node (A) at (-1.2, -6.5) {PC = };
        \node (A) at (0, -6.5) {$-0.160$};
        \node (A) at (3, -6.5) {$-0.069$};
        \node (A) at (6.2, -6.5) {$0.021$};
        \node (A) at (9.2, -6.5) {$0.110$};
        \node (A) at (12.2, -6.5) {$0.202$};
        \node (A) at (-1.55, -7) {QS error = };
        \node (A) at (0, -7) {2.359e-02};
        \node (A) at (3, -7) {5.141e-03};
        \node (A) at (6.2, -7) {1.067e-03};
        \node (A) at (9.2, -7) {3.434e-04};
        \node (A) at (12.2, -7) {3.777e-04}; 

    \end{tikzpicture}
    }
    \caption{Continuum of devices in QUASR.  The first and second row correspond to the left and right cluster in Figure \ref{fig:pca_newQH}, respectively.  The numbers below these devices are their corresponding principal component values, and quasisymmetry errors.
    The color on the magnetic surfaces corresponds to the local field strength.
    }
    \label{fig:continuum}
\end{figure}

We observe that the two clusters appear to lie on separate one-dimensional linear manifolds.  Subselecting all the devices in the first cluster where $\text{PC1} < 0$, we do a second PCA on the full feature vector with $D=663$ entries, and retaining only a single principal component.  It appears that the devices are well approximated by a one-dimensional linear manifold, because $\lambda_1 / (\lambda_1 + \hdots + \lambda_{663}) = 0.916$.  
The same thing can be done for devices in the second cluster, i.e. where $\text{PC1}>0$.
Again, the devices here are well approximated by a one-dimensional linear manifold because $\lambda_1 / (\lambda_1 + \hdots + \lambda_{663}) = 0.876$.  
After these separate one-dimensional PCAs, each device from the first cluster now has a single coordinate called PC, which lies on the interval $[\min(\text{PC}), \max(\text{PC})]$.  We can easily visualize the progression of the device geometries along these one-dimensional manifold by sampling devices uniformly according to the principal component label, i.e., devices with a principal component label that is approximately $\min(\text{PC1}) + i(\max(\text{PC})-\min(\text{PC}))/N$ for $i = 0, \hdots, N$.  In Figure \ref{fig:pca_newQH}, we have projected these one-dimensional manifolds onto the plane, indicated by two red dashed lines.  The progression of devices along these lines in both clusters is shown in Figure \ref{fig:continuum}, where the transition between the endpoint geometries appears smooth.  This is evidence that the devices in each cluster in the full $D=663$ dimensional feature space lie on simple one-dimensional continua. 
\par
The difference between the two clusters can in this way also be made clearer. We said before that $\mathbf{d}^*_1$ appeared to control the elongation of cross sections, with cluster 2 characterized by an increased value for it. However, from Figure~\ref{fig:continuum}, there is a key geometric difference between the clusters that explains this observation and which has to do with the difference in twisting of the devices. At the sharpest corners in the $RZ$ projection, cluster 1 devices have a magnetic axis mainly lying on the plane, while cluster 2 devices have their axes coming off the plane. This difference in bending and twisting is key in controlling rotational transform and other properties, and is worth further exploration in the future.

\subsection{Quasihelical symmetry  ($n_{\text{fp}}=4, \iota=1.1$)}
\begin{figure}
    \centering
    \includegraphics[width=\linewidth]{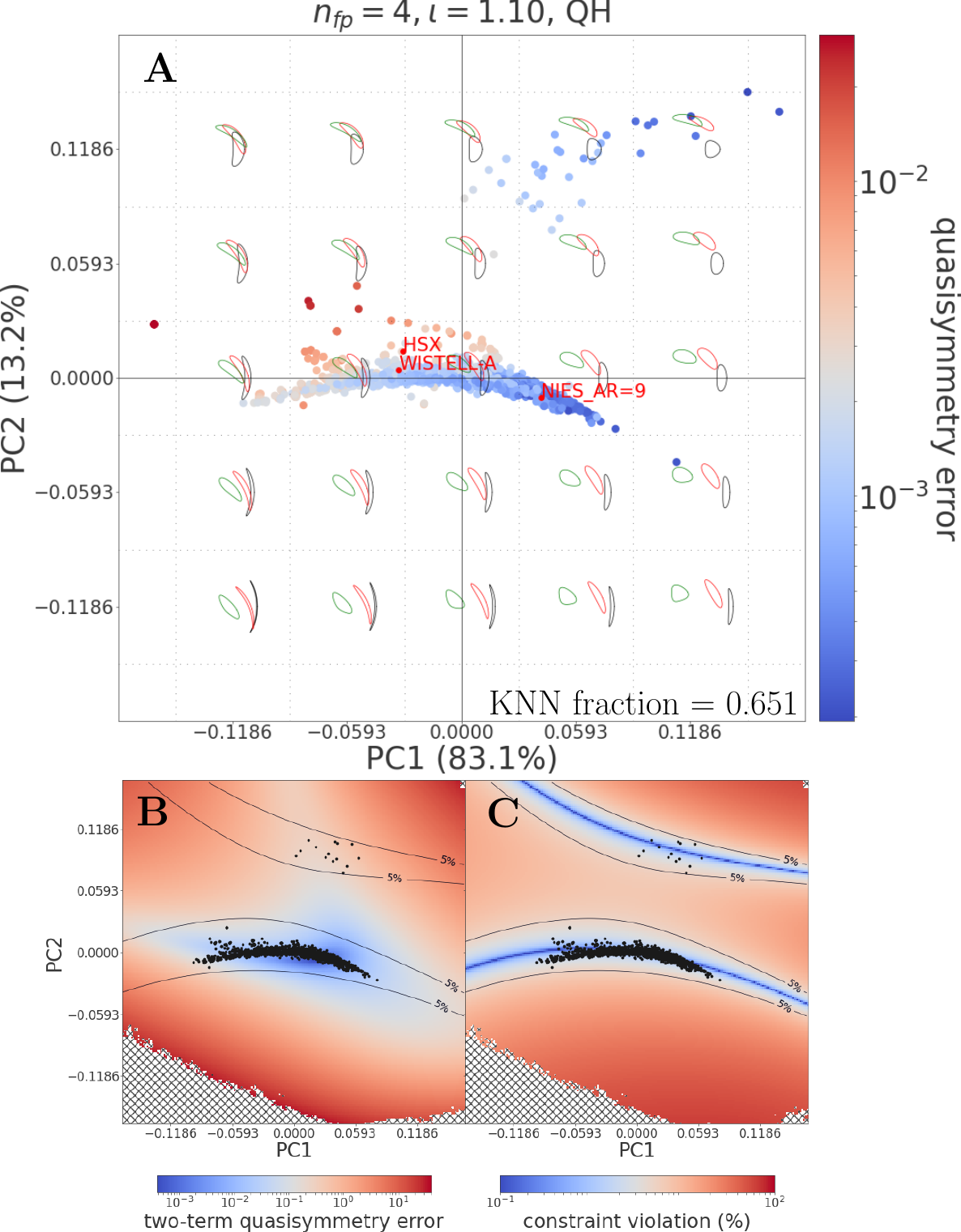}

    \caption{Panel A contains a PCA of quasihelically symmetric devices in QUASR when $\iota=1.1$ and $n_{\text{fp}}=4$, where devices in QUASR with two dominant PC components are plotted.
    Panels B, and C show respectively the two-term quasisymmetry error from \citep{PhysRevLett.101.145003,lp} computed using VMEC and relative $\iota$ constraint violation ($|\iota-1.1|/1.1$) on the PC subspace.  The black isolines correspond to coordinates where the constraint violation is precisely 5\% and the black dots are devices in panel A that have only two dominant principal components.
    The region on the PC plane where the two-term quasisymmetry error could not be evaluated or VMEC did not converge, notably on the bottom left of the panels, is hatched.}
    \label{fig:qh_map4}
\end{figure}

We look at another region of the QH design space that has been explored in the literature ($n_{\text{fp}}=4$ and $\iota=1.1$) in Figure \ref{fig:qh_map4}, where there are two clusters of devices once again, in this case with the largest centered at the origin, the second lying above it.
The \texttt{HSX}, \texttt{WISTELL-A}, \texttt{NIES\_AR=9} devices are given within the main cluster for reference. There is in this case a large disparity between clusters, with the top one being scarcely populated.
\par
To try to make sense of this particular clustering of configurations, we consider constructing and characterising vacuum magnetic fields for the whole PC plane. Each pair of coordinates in such a plane corresponds to one particular stellarator geometry, whose equilibrium we may then solve using VMEC. This allows us to understand the appearence of the cluster, as well as opening the door to analysing other configuration properties within such a space.  We present a measure of QS quality over this plane in Figure~\ref{fig:qh_map4}B (in particular a plot of the two-term quasisymmetry residual from \citep{lp, PhysRevLett.101.145003}, also $f_C$ in \cite{rodriguez2022measures}), and overlay as black dots the QUASR devices in the clusters above with two dominant PC components. As expected, two agglomerations of devices in QUASR lie in local QS minima, providing external validation that the devices in QUASR were properly optimized for quasisymmetry. However, it fails to properly describe the separation into the two clusters, for which one must consider the additional design constraint that the average rotational transform is $\iota=1.1$. Figure \ref{fig:qh_map4}C shows the relative violation of this constraint of rotational transform on the plane; \textit{i.e.}, $|\iota-1.1|/1.1$. 
It appears that there are two separate manifolds where the $\iota$ design constraint is accurately satisfied. Since the devices in QUASR were designed to satisfy these constraints to within $0.1\%$ accuracy, the configuration clusters lie in the intersection of these delineated regions and Figure~\ref{fig:qh_map4}B. This is not exact, though, especially because in reality devices have more than only two principal components, and are not located precisely on the PC1-PC2 plane. 
\par
The topology of the data appears thus to be governed by the interplay of the rotational transform design constraint and the valleys of the quasisymmetry objective. This perspective on the origin of the clusters also explains why the top cluster is more scarcely populated, as it has less accurate quasisymmetry than the bottom one and hence corresponds to a shallower optimization well. 
This phenomenon could also explain the clusters in Section~\ref{sec:qh_nfp4_iota2p3}, but VMEC failing in a large portion of the PC1-PC2 plane prevents any conclusion.

\subsection{Quasihelical symmetry  ($n_{\text{fp}}=3, \iota=1.1$)}

\begin{figure}
    \centering

    \includegraphics[width=\textwidth]{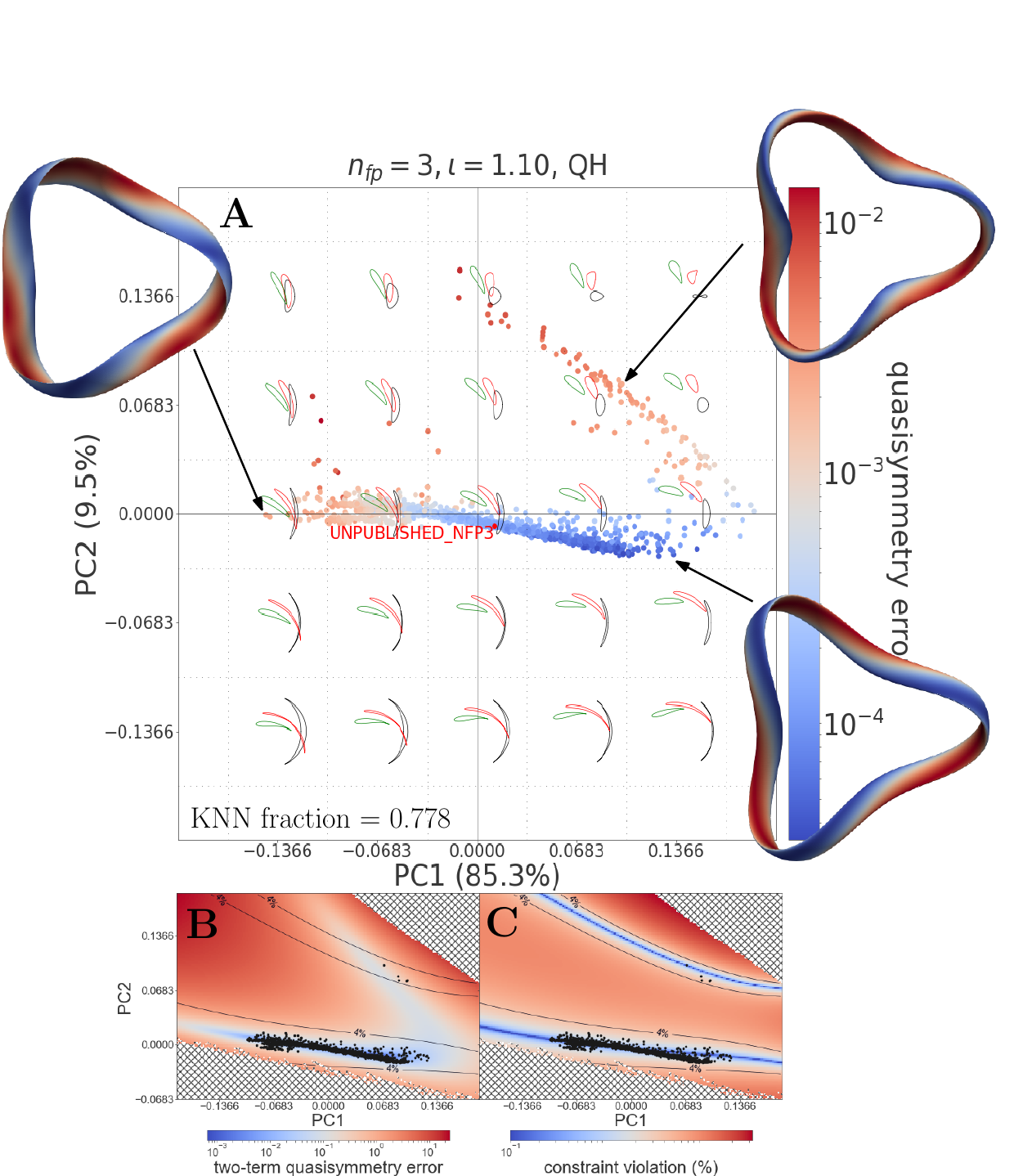}
    \caption{Panel A contains a PCA of quasihelically symmetric devices in QUASR, along with an unpublished device from \citep{kappel_magnetic_2024}, indicated with red text.      The color on the magnetic surfaces corresponds to the local field strength.
    Panels B, and C show respectively the two-term quasisymmetry error from \citep{PhysRevLett.101.145003,lp} computed using VMEC and relative $\iota$ constraint violation ($|\iota-1.1|/1.1$) on the PC subspace.  The black isolines correspond to coordinates where the constraint violation is precisely 4\% and the black dots are devices in panel A that have only two dominant principal components.
    The region on the PC plane where the two-term quasisymmetry error could not be evaluated or VMEC did not converge is hatched.
    }
    \label{fig:pca_qh1}
\end{figure}

In Figure \ref{fig:pca_qh1}A, we plot the landscape of devices for a lower rotational transform $\iota=1.1$ and $n_{\text{fp}}=3$, where it appears that little information was lost in the projection.
We find the devices in QUASR here are close to an unpublished device from \citep{kappel_magnetic_2024}.
The KNN fraction is quite high, 0.778.
In contrast to the PCA in section \ref{sec:qh_nfp4_iota2p3}, both principal components appear to modify the devices' elongation and the positioning of the cross sections.
We note that here, there does not appear to be two separate clusters of devices and the devices form a V-shaped continuum.  
Figures~\ref{fig:pca_qh1}B, C contain the two-term quasisymmetry error and $\iota$ constraint violation. The QUASR devices appear to follow the QS well throughout the plane (even suggesting the V-shape), but this clearly clashes with the rotational transform constraint. It does not in the lower portion of the cluster, which explains the large number of field cases found there. But it does in the middle portion.
\par
Behind this apparent contradiction lies an important observation we emphasised before. Figures~\ref{fig:pca_qh1}B, C correspond to configurations with only two principal components. But the QUASR devices truly live in a higher dimensional space. In particular, the configurations in the elbow of the V-shape cluster do, which are in fact the devices that do not fit with the interpretation of Figures~\ref{fig:pca_qh1}B, C.

\begin{figure}
    \centering
    \resizebox{\textwidth}{!}{
    \begin{tikzpicture}
        \node (A) at (0, -8.5) {\includegraphics[width=0.175\textwidth, trim=300 50 280 50, clip]{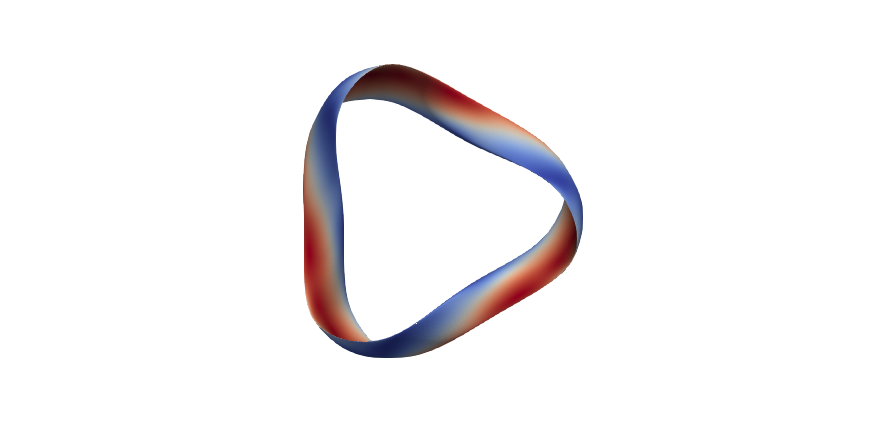}};
        \node (B) at (3.2,-8.5) {\includegraphics[width=0.175\textwidth, trim=300 50 280 50, clip]{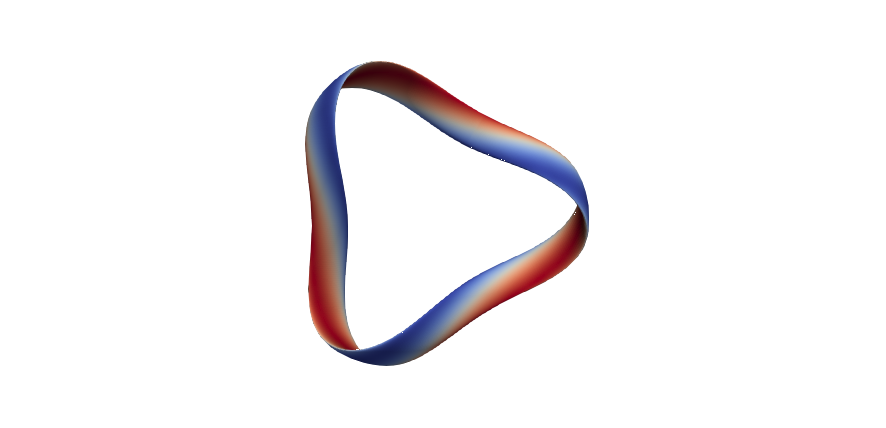}};
        \node (C) at (6.2,-8.5) {\includegraphics[width=0.175\textwidth, trim=300 50 280 50, clip]{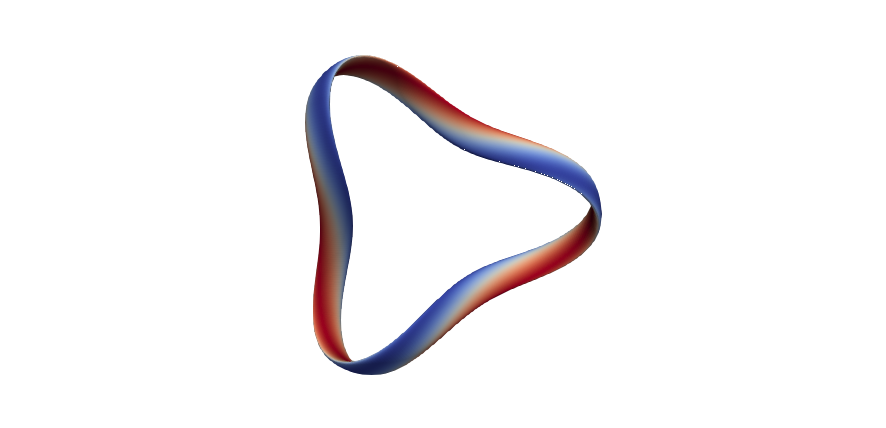}};
        \node (D) at (9.2,-8.5) {\includegraphics[width=0.175\textwidth, trim=300 50 280 50, clip]{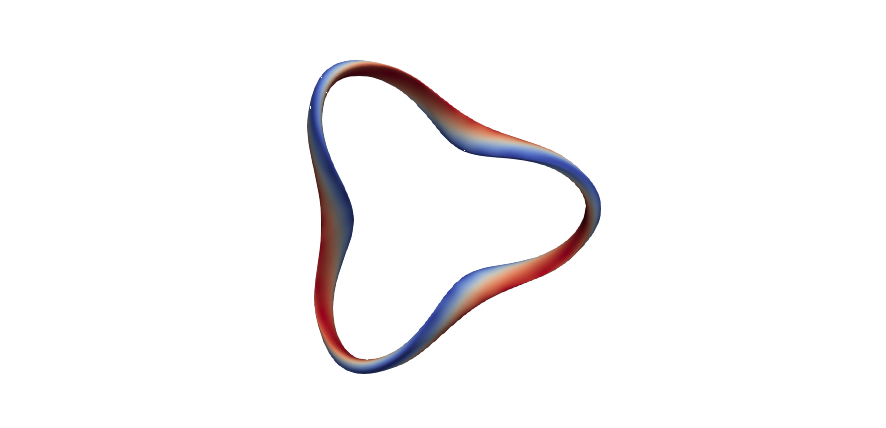}};
        \node (E) at (12.2,-8.5) {\includegraphics[width=0.175\textwidth, trim=300 50 280 50, clip]{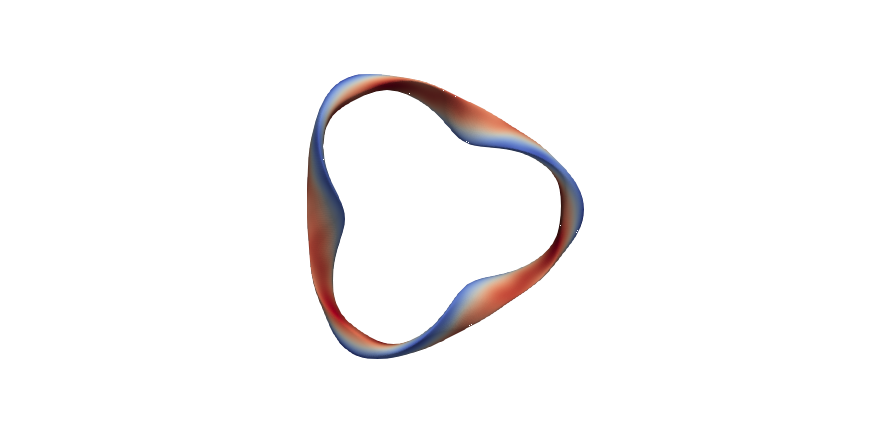}};

        \node (A) at (-1.6, -10.25) {Isomap};
        \node (A) at (-1.6, -10.7) {coord.};
        \node (A) at (-0.85, -10.5) {=};
        \node (A) at (0.1, -10.5) {$-0.172$};
        \node (A) at (3, -10.5) {$-0.030$};
        \node (A) at (6.2, -10.5) {$0.111$};
        \node (A) at (9.2, -10.5) {$0.253$};
        \node (A) at (12.2, -10.5) {$0.394$};

        \node (A) at (-1.6, -11.25) {QS error = };
        \node (A) at (0, -11.25) {2.324e-03};
        \node (A) at (3, -11.25) {2.119e-04};
        \node (A) at (6.2, -11.25) {8.735e-05};
        \node (A) at (9.2, -11.25) {2.315e-03};
        \node (A) at (12.2, -11.25) {1.077e-02};  

    \end{tikzpicture}
    }
    \caption{The nonlinear manifold in Figure \ref{fig:pca_qh1} can be parametrized with a single coordinate using the isomap algorithm.  The row of devices are uniformly sampled from the isomap embedding, and the numbers below these devices corresponds to the isomap coordinate and the quality of quasisymmetry. The color on the magnetic surfaces corresponds to the local field strength}
    \label{fig:isomap}
\end{figure}

We may nevertheless attempt to describe the cluster in this scenario as a one-dimensional continuum. However, a PCA is inappropriate here as clearly the manifold is nonlinear.
Instead, we use the isomap algorithm \citep{tenenbaum2000global} to visualize the progression of the device geometries along the manifold.  The isomap algorithm works by generating a connectivity graph between data points using the $k$-nearest neighbors of each node.  Then, pairwise distances between points are computed by traversing the connectivity graph.  The multidimensional scaling algorithm \citep{kruskal1964nonmetric} attempts to find a lower dimensional representation that preserves these pairwise distances. 
Our aim is to find a one-dimensional labeling of the devices so that the devices can be ordered, similar to Section \ref{sec:qh_nfp4_iota2p3}.  Applying the isomap algorithm on the full $D=663$ dimensional data points, we obtain the embedding shown in Figure \ref{fig:isomap}, i.e., each device is given a one-dimensional coordinate.  Uniformly sampling this coordinate, the progression of devices in the bottom row of the figure is obtained. 

\subsection{Quasihelical symmetry  ($n_{\text{fp}}=5, \iota=2.5$)}

\begin{figure}
\centering
\includegraphics[width=\textwidth, trim=15 45 30 30, clip]{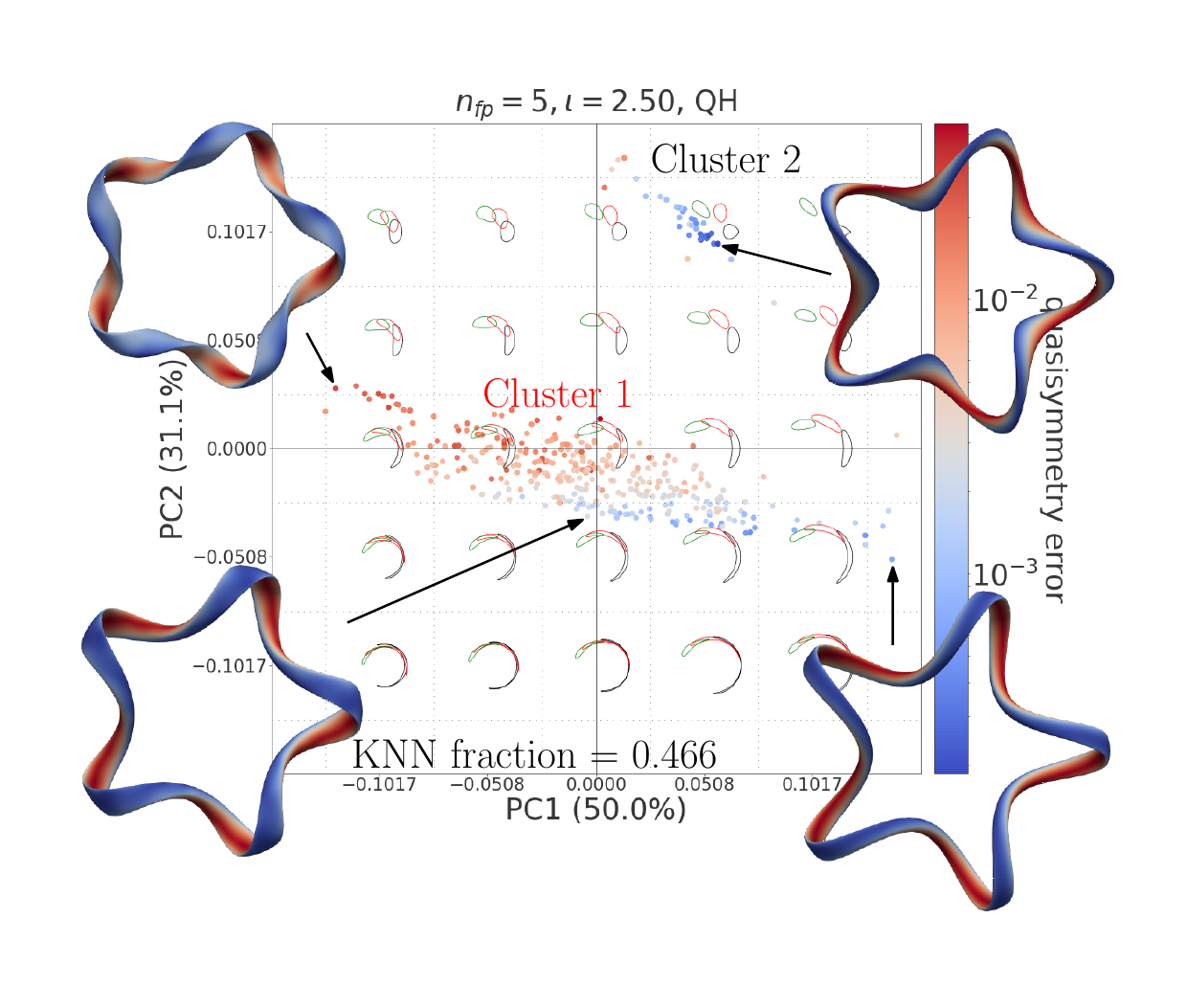}

    \caption{PCA of quasihelically symmetric devices in QUASR when $\iota=2.5$ and $n_{\text{fp}}=5$.  Two clusters are labeled and devices from them are plotted.
        The color on the magnetic surfaces corresponds to the local field strength.
    }
    \label{fig:pca_qh2}
\end{figure}
In this section, we show the two-dimensional landscape of devices with a rotational transform $\iota=2.50$ and $n_{\text{fp}}=5$ in  Figure \ref{fig:pca_qh2}.
The first principal component appears to shift the relative positions of the cross sections, while the second principal component appears to act on the elongation of the device (similar to the $n_\mathrm{fp}=4$ before).
There appears to be two separate clusters of devices, where the second cluster contains the most quasisymmetric devices (again there seems to be a parallel to the $n_\mathrm{fp}=4$ scenario, even on the geometric behavior within and between clusters).
\par
However, a non-negligible amount of information was lost in this projection, as the sum of the PC ratios is just over 80\%.  Similarly, the KNN fraction is just over 0.45.
\begin{figure}
    \centering
    \resizebox{\textwidth}{!}{
    \begin{tikzpicture}
    \node (A) at (0, 0) {\includegraphics[width=0.6\linewidth]{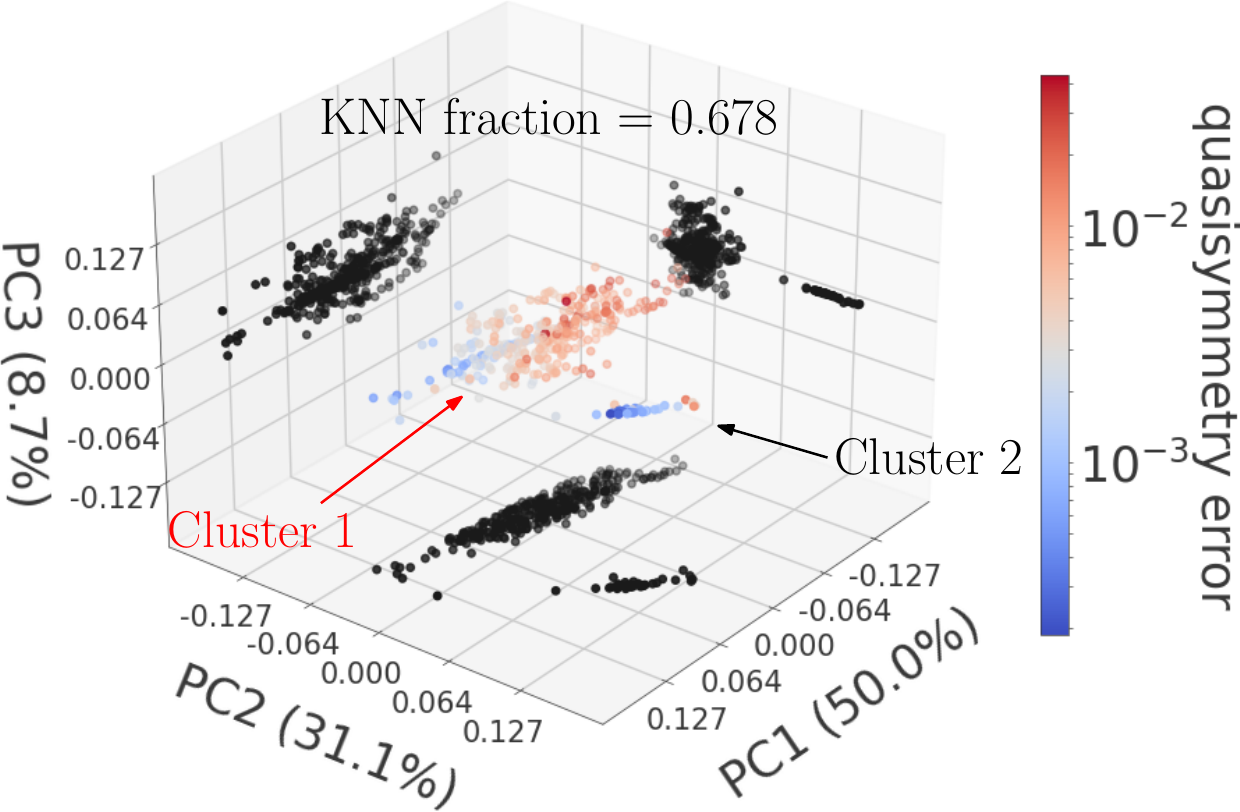}};
    \node (B) at (-7, 0) {\includegraphics[width=0.33\linewidth]{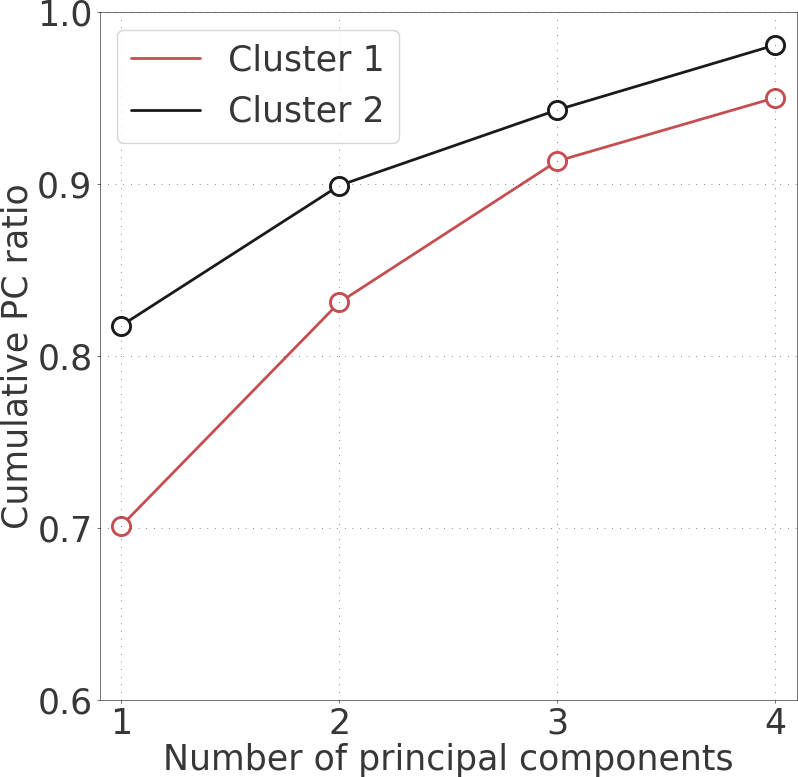}};
    \end{tikzpicture}
    }
    \caption{On the left, the cumulative PC ratio with increasing number of principal components used to represent the two clusters from Figure \ref{fig:pca_qh2}.  On the right, PCA of quasihelically symmetric devices in QUASR when $\iota=2.5$ and $n_{\text{fp}}=5$, similar to Figure \ref{fig:pca_qh2} but with a third principal component.   Two-dimensional projections of the data on the PC1-PC2, PC1-PC3, and PC2-PC3 planes are also provided in black to make the three-dimensional nature of the data more easily understood.}
    \label{fig:qh_map3}
\end{figure}
To provide a more accurate representation of the data, we must augment Figure \ref{fig:pca_qh2} with a third principal component in Figure \ref{fig:qh_map3}.  Two-dimensional projections of the data on the PC1-PC2, PC1-PC3, and PC2-PC3 planes are also provided to make the three-dimensional nature of the data more easily understood.
Our projection error measures immediately improve as the sum of the PC ratios is approximately 90\%, and the KNN fraction increases to 0.678.
\par
The third PC component evidences that the clusters are rather different: the first cluster is a three-dimensional cloud, while the second cluster appears to have very little variation in the third PC.
The higher dimensionality of cluster 1 can be quantified by observing the cumulative PC ratio as the number of principal components used to represent cluster 1 and 2 on the left of Figure \ref{fig:qh_map3}.  It takes three PCs for the cumulative PC ratio of cluster 1 to pass 0.9, while it takes only two for cluster 2.

\subsection{Quasiaxisymmetry}
\begin{figure}
    \centering
    \includegraphics[width=\linewidth]{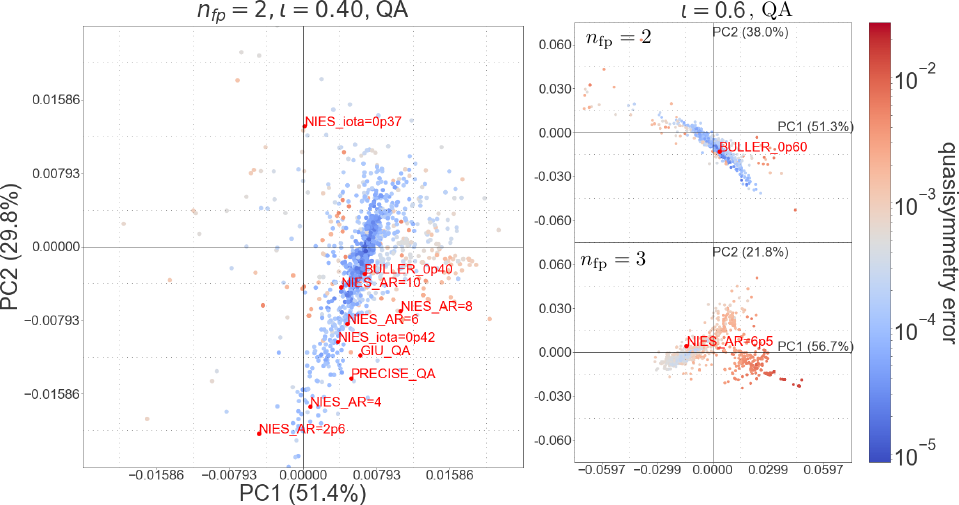}
    \caption{PCA of quasiaxisymmetric devices in QUASR along with devices in the literature, indicated with red text.}
    \label{fig:pca_qa12}
\end{figure}

For completeness, on the left of Figure \ref{fig:pca_qa12}, we complete a PCA on quasiaxisymmetric devices in QUASR with $n_{\text{fp}}=2$ and $\iota=0.4$.  The cumulative PC ratio is just over $80\%$, revealing that a non-negligible amount of information is lost in this projection, and it might be useful to include a third principal component in the visualization.  
Given the two principal axes, we project devices outside QUASR onto the figure and label them.  This area of the parameter space is well explored.
In Figure \ref{fig:pca_qa12}, we do a PCA for QA devices, subselected based on their rotational transform ($\iota= 0.6$) and number of field periods ($n_{\text{fp}}=2,3$).
Labeled devices from the literature appear similar to the devices in QUASR.  We do not provide cross sections of hypothetical devices on the linear manifold as we did in previous sections because, visually, the cross sections do not differ significantly from each other.

The analysis from section \ref{sec:nae_landscape} reveals that the region of QA devices with favorable quasisymmetry is small, and moreover it shrinks with increasing shaping of the device.  This bears out in the PCA analysis here too, as we find that there is not much geometric diversity of the devices at the aspect ratio studied.  Notably, the Landreman-Paul QA device (\texttt{PRECISE\_QA}) is close on the plane to the device called \texttt{GIU\_QA}.  This device was obtained by providing the \texttt{PRECISE\_QA} device as an initial guess to a direct coil optimization algorithm.  The quality of quasisymmetry in these two devices differs by almost an order of magnitude, despite the geometrical proximity on the plot.

\section{Discussion and conclusion}
The main goal of this paper was to introduce a new data set of optimized stellarators, and provide tools to navigate the contents of it.
We start by highlighting a few devices in QUASR first, and then move to analysing the data-set. First, we frame QUASR within the near-axis framework and its predicted landscape of quasisymmetric configurations. We find close correspondence of devices in QUASR to regions of favorable near-axis quasisymmetry.
However, this approach does not apply to the full set of devices in QUASR given the difference in assumptions that went into their construction, most notably constraints on elongation and simplicity of the shaping. A more model agnostic procedure to analyze the data is then presented in the form of a PCA.
On the subsets of the data set that we have explored in this work, we find that the first and second principal components act predominantly on the elongation (and different field twisting), and relative positions of cross sections of the magnetic surfaces. These two dimensions are able to capture the devices in the data set to a large degree.
For some subsets of the data, we observe clear clusters.  
Within the clusters, we find evidence that the devices form a continuum whose shape follows that of the design constraint manifold.

\begin{figure}
    \centering
    \includegraphics[width=\linewidth]{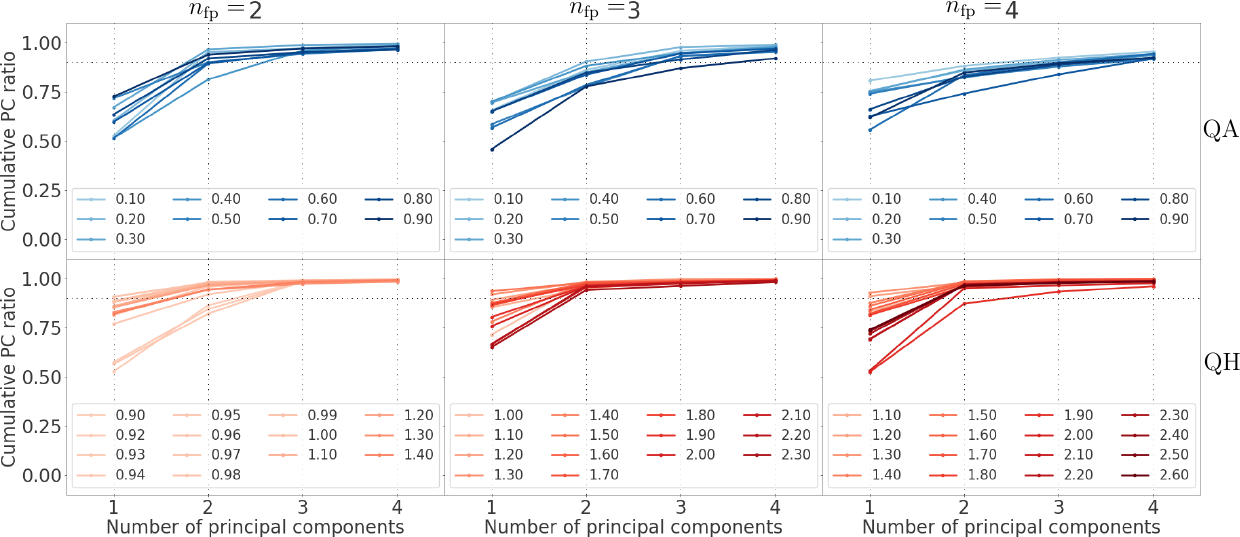}
    \caption{The cumulative PC ratio of subsets of the data for various combinations of $n_{\text{fp}}, \iota$ and helicity. The first and second rows correspond to QA and QH devices, respectively.  The columns correspond to particular values of $n_{\text{fp}}$, and the curves correspond to different values of $\iota$.
    The horizontal dashed line corresponds to a cumulative PC ratio of 0.9.}
    \label{fig:complexity}
\end{figure}

For the sets presented,  two to three principal components appear sufficient for visualization. This is true of even more subsets throughout the QUASR database, as illustrated in Figure \ref{fig:complexity}, where we compute a PCA for various combinations of $n_{\text{fp}}$ and $\iota$ for QA and QH devices. It is possible to attain a cumulative PC ratio of 0.9 consistently.
This illustrates that much of the variation between the devices can be characterized on a low-dimensional, linear manifold. This points to the lower dimensional nature of the subset of quasisymmetric configurations in the larger space of stellarators.  This echoes the analysis in \citep{sengupta2023periodic}, where it was revealed that quasisymmetry imposes a lower-dimensionality on the field strength.  We would like to highlight though that this lower-dimensionality does not necessarily extend to the \textit{coil} design space, which is known to be complex with multiple minima \citep{quasr1, Wechsung_2022}.  This is because there are many, vastly different coil sets that produce approximately the same magnetic field \citep{Landreman_2017}.  This highlights that some sort of global search or a problem reformulation is indeed necessary for coil design.

Whether the information lost in the dimensionality reduction is significant depends on the application.  For the purposes of visualization in this work, a cumulative PC ratio of 90\% appeared satisfactory.  But in other applications, this might not be the case.
Some useful extensions of this include using the very simple, linear formulas learned during the PCA to generate warm starts for optimization.
Alternatively, one might attempt optimizing in the PC subspace.
These directions are not explored in this paper.

\section{Data availability}
The latest version of the QUASR data set can be found at \url{https://zenodo.org/doi/10.5281/zenodo.10050655}.
It can be visually explored at \url{quasr.flatironinstitute.org}.

\section{Acknowledgements}
The authors would like to thank Flatiron's Scientific Computing Core (SCC).  We would also like to thank Marsha Berger, Rogerio Jorge, Matt Landreman, Elizabeth Paul, Manas Rachh, and Lawrence Saul for helpful discussions. E. R. was supported by a grant of the Alexander-von-Humboldt-Stiftung, Bonn, Germany, through a postdoctoral research fellowship.

\appendix

\appendix
\section{Fitting the second-order near-axis model to the devices}\label{sec:fitting}
In this Appendix we describe how a near-axis model to second order is constructed for a given field from the QUASR database.
\par
The geometry of the field's magnetic axis can be directly constructed from the given coils, but an additional three parameters are needed that describe the second order near-axis model in vacuum: $B_0$, $\etabar$ and $B_{2c}$.  In the neighborhood of the magnetic axis, the field strength $B = \|\mathbf B\|$ of the near-axis model satisfies
$$
B(r, \varphi, \theta) = B_0 + rB_0\etabar\cos(\theta - N\varphi)  + r^2[B_{20}(\varphi) + B_{2c}\cos(2(\theta - N\varphi))] + \hdots,
$$
where stellarator symmetry is assumed, $\varphi, \theta$ are Boozer angles, and $r$ is the radial distance off-axis. 
There are a multitude of methods to estimate these near-axis parameters in the field generated by the coils.  
For instance, one could evaluate the field strength on a three-dimensional tensor-product grid $(r_i, \varphi_j, \theta_k)$ in the neighborhood of the magnetic axis and solve for the values of $B_0, \etabar, B_{2c}, B_{20}(\varphi_j)$ in a least-squares sense.
Another possibility is to fit the predicted cross-sections of the near-axis magnetic surfaces to the true ones \citep{Landreman_2019}.
These techniques both have the difficulty that one must use information off axis, and the quality of the fit might be sensitive to the choice of how far off-axis one goes.
In order to make the fitting procedure independent of this user-defined choice, we solve three minimization problems sequentially:
\begin{equation}
\begin{aligned}
B_0^* &= \argmin_{B_0} \int_{0}^{2\pi} \|\mathbf B_{\text{coils}} - \mathbf B_{\text{NAE}}(B_0)\|^2 ~d\phi ,\\
\etabar^* &= \argmin_{\etabar} \int_{0}^{2\pi} \|\nabla \mathbf B_{\text{coils}} - \nabla \mathbf B_{\text{NAE}}(B_0^*, \etabar) \|^2 ~d\phi ,\\
B_{2c}^* &= \argmin_{B_{2c}} \int_{0}^{2\pi} \|\nabla \nabla \mathbf B_{\text{coils}} - \nabla \nabla \mathbf B_{\text{NAE}}(B_0^*, \etabar^*, B_{2c}) \|^2 ~d\phi,
\end{aligned}
\end{equation}
where $\phi$ is the standard cylindrical angle.
Since the magnetic field is generated by a set of coils, we can compute the magnetic field as well as its first and second derivatives using the Biot-Savart law, which we have direct access to in SIMSOPT.
As a result, we can find the near-axis parameters that most closely reproduce the coils' magnetic field and its derivatives on axis.
Note that the near-axis model assumes that there is perfect quasisymmetry to first order.
While a good approximation of quasisymmetry is attainable in the volume, the devices do not present perfect near-axis quasisymmetry, and we may not always be able to obtain be a good fit of the coils' magnetic field.  

\begin{figure}
    \centering
    \includegraphics[width=\linewidth]{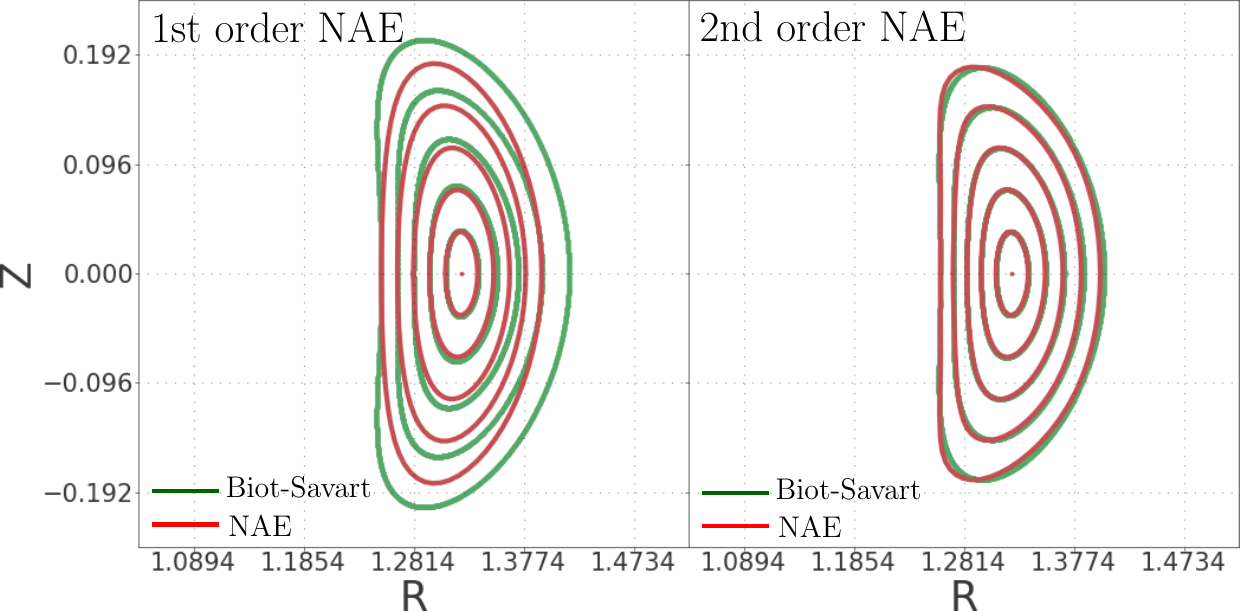}
    \caption{Poincar{\'e} plot (green) and cross sections of surfaces generated by the near-axis expansion (red) fit to a device in QUASR. The left panel shows the first order NAE surfaces, while the right panel shows the second order NAE surfaces.  The outermost surfaces here have aspect ratio $\sim 9$.  The field lines are generated from starting points along the inboard side with $Z=0$.  The Biot-Savart Poincar{\'e} sections are different on the left and right panels because the field lines are initialized from different starting points.}
    \label{fig:nae_vs_full}
\end{figure}

In Figure \ref{fig:nae_vs_full}, we provide a Poincar{\'e} plot alongside cross sections of surfaces generated by a first and second order near-axis expansion.  The NAE model was fit using the method described in this section, revealing that the second order NAE is capable of describing the magnetic surfaces in the neighborhood of the magnetic axis much more faithfully than the first order one.

\section{PCA example for Gaussian processes}  \label{app:gaussians}
To gain some intuition on the quality measures of cumulative PC ratios and KNN fraction introduced in Section~\ref{sec:PCA}, we apply PCA to data generated from three Gaussians in this Appendix. These Gaussians are defined to have mean and covariance matrices,
\begin{alignat*}{5}
    \mu_1 &= \begin{pmatrix}
        -6 \\
        0\\
        0
    \end{pmatrix} &&~\Sigma_1 = Q_1^T\begin{pmatrix}
        1 & 0 &0 \\
        0 & 10^{-1} & 0\\
        0 & 0 & 10^{-1}
    \end{pmatrix}Q_1\\
    \mu_2 &=\begin{pmatrix}
        6 \\
        -6\\
        -6
    \end{pmatrix} , &&~\Sigma_2 = Q_2^T\begin{pmatrix}
        1 & 0 &0 \\
        0 & 1 & 0\\
        0 & 0 & 10^{-1}
    \end{pmatrix}Q_2, \\
\mu_3 &=\begin{pmatrix}
        6 \\
        0\\
        0
    \end{pmatrix} ,&&~\Sigma_3 = \begin{pmatrix}
        1 & 0 &0 \\
        0 & 1 & 0\\
        0 & 0 & 1
    \end{pmatrix},
\end{alignat*}
where $Q_1, Q_2$ are randomly generated rotation matrices.   Individually, the Gaussians are well approximated by linear manifolds of dimension one, two, and three, respectively.  Samples from these distributions are shown on the left of Figure \ref{fig:gaussians} in the main text, while a two-dimensional projection of the data is shown on the right.
\par
In 2D, the cumulative PC ratio is quite high 0.986, revealing that the majority of variation in the data are captured by two principal components.   Despite this favorable quality measure and such a simple data set, the KNN fraction is approximately 0.4, i.e.,  $\approx$40\% of the 10 nearest neighbors in 3D remain so on the 2D projection on average. 
\begin{table}
    \centering
    \begin{tabular}{ccccc}
      & \# PC components & 1 & 2  & 3\\

        \cline{2-5}
       
     \multirow{3}{5em}{cumulative PC ratios}   & cluster 1  & 0.828 & 0.916 & 1.000 \\
      & cluster 2  & 0.501 & 0.953 & 1.000\\
      & cluster 3  & 0.361 & 0.712 & 1.000\\

       \cline{2-5}

    \multirow{3}{5em}{KNN fraction} & cluster 1  & 0.216 & 0.468 & 1.000 \\
      & cluster 2  & 0.140 & 0.614 & 1.000\\
      & cluster 3  & 0.095 & 0.352 & 1.000\\
    \end{tabular}
    \caption{The top three rows are the cumulative PC ratio and the bottom three rows are the KNN fraction  with respect to the number of PCs for the clusters in Figure \ref{fig:gaussians}.  }
    \label{tab:cumulative}
\end{table}

 In Table \ref{tab:cumulative}, we provide the cumulative PC ratio as a function of the number of PCs.  In order for the PC ratio to attain 0.9, each cluster requires a different number of PCs.  This hints that cluster 1 is well approximated on a line, cluster 2 on a plane, and cluster 3 on a volume.  In fact, since cluster 3 is genuinely three-dimensional, dimensionality reduction is not possible without significant loss of information.
 The KNN fractions increase with the number of PCs, with a value greater than 0.4 appearing to be a lower bound on a reasonable lower-dimensional representation.

\section{Reference QS configurations from the literature} \label{app:configs}
The QS configurations from the literature referenced throughout the paper are summarized in Table \ref{tab:devices}. These devices were all optimized for QS at various aspect ratios, and target rotational transforms.  Moreover, some devices were designed to have a target mean rotational transform \citep{lp, surfaceopt1}, while others had a target edge rotational transform \citep{nies2024exploration}.
\begin{table}
    \centering
    \begin{tabular}{ccccccc}
        Label  & $n_{\text{fp}}$ & aspect ratio & $\iota$  & QS type&  Figure \# & Reference \\
        \texttt{BULLER\_0p40} & 2 & 6.00 &  0.40 (m) & QA & \ref{fig:pca_qa12} & \cite{buller_family_2024}\\
        \texttt{GIU\_QA} & 2 & 6.00 & 0.42 (m)& QA& \ref{fig:pca_qa12} & \cite{surfaceopt1}\\
        \texttt{NIES\_iota=0p37} & 2 & 6.00 & 0.37 (e) & QA& \ref{fig:pca_qa12} & \cite{nies2024exploration}\\
        \texttt{NIES\_iota=0p42} & 2 & 6.00 &  0.42 (e) & QA& \ref{fig:pca_qa12} & \cite{nies2024exploration}\\
        \texttt{NIES\_AR=10} & 2 & 10.0 & 0.42 (e) & QA& \ref{fig:pca_qa12} & \cite{nies2024exploration}\\
        \texttt{NIES\_AR=8} & 2 & 8.00 & 0.42 (e) & QA& \ref{fig:pca_qa12} & \cite{nies2024exploration}\\
        \texttt{NIES\_AR=6} & 2 & 6.00 & 0.42 (e) & QA& \ref{fig:pca_qa12} & \cite{nies2024exploration}\\
        \texttt{NIES\_AR=4} & 2 & 4.00 & 0.42 (e) & QA& \ref{fig:pca_qa12} & \cite{nies2024exploration}\\
        \texttt{NIES\_AR=2p6} & 2 & 2.60 & 0.42 (e) & QA&  \ref{fig:pca_qa12} & \cite{nies2024exploration}\\
        \texttt{PRECISE\_QA} & 2 & 6.00 & 0.42 (m) & QA & \ref{fig:pca_qa12} & \cite{lp}\\
        \texttt{BULLER\_0p60} & 2 & 6.00 &  0.60 (m) & QA & \ref{fig:pca_qa12} & \cite{buller_family_2024}\\
        \texttt{NIES\_AR=6p5} & 3 & 6.50 & 0.42 (e) & QA & \ref{fig:pca_qa12} & \cite{nies2024exploration}\\
        \texttt{UNPUBLISHED\_NFP3} & 3 & 6.00 &  1.06 (m) & QH  & \ref{fig:pca_qh2} & \cite{kappel_magnetic_2024}\\
        \texttt{NEW\_QH}  & 4 & 13.5 & 2.29 (m) & QH & \ref{fig:pca_newQH} & \cite{rodriguez2023constructing}\\
        \texttt{HSX} (vacuum)  & 4 & 10.0 & 1.06 (m) & QH & \ref{fig:qh_map4}  &  \cite{anderson1995helically}\\
        \texttt{WISTELL-A}  & 4 & 6.72 & 1.1 (m)  & QH & \ref{fig:qh_map4} & \cite{bader2020advancing} \\
        \texttt{NIES\_AR=9}  & 4 & 9.00 & 1.3 (e) & QH & \ref{fig:qh_map4} & \cite{nies2024exploration}\\
    \end{tabular}
    \caption{Labeling of devices from the literature, the figure number on which they appear in this manuscript, along with the value of $n_{\text{fp}}$, type of QS targeted, aspect ratio, and rotational transform.  We quote either the mean rotational transform, or the edge rotational transform.  
    These separate cases are highlighted with an `m' or `e' in parentheses.
    }
    \label{tab:devices}
\end{table}
\par
To make a direct comparison of these to the devices in the QUASR dataset, we must construct their corresponding feature vectors using the following procedure. 
First, we take the geometry of the device's outermost surface and use VMEC \citep{VMEC} to determine the vacuum magnetic field on the entire toroidal volume.  We take the surface in the device that has aspect ratio closest to 10 for QA and 12 for QH.  The Boozer coordinates on this surface are computed using \texttt{BOOZ\_XFORM} \citep{sanchez2000ballooning}.  Finally, we compute the Fourier harmonics of this surface parametrized in Boozer coordinates. We may then treat them as we did with QUASR objects, and when necessary project them to the appropriate principal component plots.  
We directly compare these configurations to the relevant ones in QUASR when they share the value of $n_{\text{fp}}$, helicity, and mean rotational transform (out to aspect ratio 10 (for QA) and 12 (for QH) to an accuracy of $\pm 0.05$). %
\par

\bibliographystyle{jpp}
\bibliography{references}

\begin{thebibliography}{60}
\expandafter\ifx\csname natexlab\endcsname\relax\def\natexlab#1{#1}\fi
\def\au#1{#1} \def\ed#1{#1} \def\yr#1{#1}\def\at#1{#1}\def\jt#1{\textit{#1}}
  \def\bt#1{#1}\def\bvol#1{\textbf{#1}} \def\vol#1{#1} \def\pg#1{#1}
  \def\publ#1{#1}\def\arxiv#1{#1}\def\org#1{#1}\def\st#1{\textit{#1}}

\bibitem[Anderson {\em et~al.\/}(1995)Anderson, Almagri, Anderson, Matthews,
  Talmadge \& Shohet]{anderson1995helically}
{\sc \au{Anderson, F. S.~B.}, \au{Almagri, A.~F.}, \au{Anderson, D.~T.},
  \au{Matthews, P.~G.}, \au{Talmadge, J.~N.} \& \au{Shohet, J.~L.}} \yr{1995}
  \at{The helically symmetric experiment, ({HSX}) goals, design and status}.
  \jt{Fusion Technology}  \bvol{27}~(3T),  \pg{273--277}.

\bibitem[Bader {\em et~al.\/}(2020)Bader, Faber, Schmitt, Anderson, Drevlak,
  Duff, Frerichs, Hegna, Kruger, Landreman {\em et~al.\/}]{bader2020advancing}
{\sc \au{Bader, A.}, \au{Faber, B.~J.}, \au{Schmitt, J.~C.}, \au{Anderson,
  D.~T.}, \au{Drevlak, M.}, \au{Duff, J.~M.}, \au{Frerichs, H.}, \au{Hegna,
  C.C.}, \au{Kruger, T.G.}, \au{Landreman, M.} \& \au{others}} \yr{2020}
  \at{Advancing the physics basis for quasi-helically symmetric stellarators}.
  \jt{Journal of Plasma Physics}  \bvol{86}~(5),  \pg{905860506}.

\bibitem[Bishop(2006)]{bishop2006pattern}
{\sc \au{Bishop, C.}} \yr{2006} {\em Pattern Recognition and Machine
  Learning\/}.  \publ{Springer}.

\bibitem[Blank(2004)]{blank2004guiding}
{\sc \au{Blank, HJ~de}} \yr{2004}  \at{Guiding center motion}.  \jt{Fusion
  science and technology}  \bvol{45}~(2T),  \pg{47--54}.

\bibitem[Bonhomme {\em et~al.\/}(2014)Bonhomme, Picq, Gaucherel \&
  Claude]{bonhomme2014momocs}
{\sc \au{Bonhomme, V.}, \au{Picq, S.}, \au{Gaucherel, C.} \& \au{Claude, J.}}
  \yr{2014}  \at{Momocs: outline analysis using {R}}.  \jt{Journal of
  Statistical Software}  \bvol{56},  \pg{1--24}.

\bibitem[Boozer(1981{\natexlab{{\em a\/}}})]{boozer1981plasma}
{\sc \au{Boozer, A.~H}} \yr{1981{\natexlab{{\em a\/}}}}  \bt{Plasma equilibrium
  with rational magnetic surfaces}. {\em Tech. Rep.\/}.  \org{Princeton Univ.}

\bibitem[Boozer(1981{\natexlab{{\em b\/}}})]{boozer1981transport}
{\sc \au{Boozer, A.~H.}} \yr{1981{\natexlab{{\em b\/}}}}  \bt{Transport and
  isomorphic equilibria}. {\em Tech. Rep.\/}.  \org{Princeton Plasma Physics
  Lab.(PPPL), Princeton, NJ (United States)}.

\bibitem[Bourlard \& Kamp(1988)]{bourlard1988auto}
{\sc \au{Bourlard, Herv{\'e}} \& \au{Kamp, Yves}} \yr{1988}
  \at{Auto-association by multilayer perceptrons and singular value
  decomposition}.  \jt{Biological cybernetics}  \bvol{59}~(4),  \pg{291--294}.

\bibitem[Buller {\em et~al.\/}(2024)Buller, Landreman, Kappel \&
  Gaur]{buller_family_2024}
{\sc \au{Buller, S.}, \au{Landreman, M.}, \au{Kappel, J.} \& \au{Gaur, R.}}
  \yr{2024} A family of quasi-axisymmetric stellarators with varied rotational
  transform. ArXiv:2401.09021 [physics].

\bibitem[Burby {\em et~al.\/}(2020)Burby, Kallinikos \& MacKay]{burby2020some}
{\sc \au{Burby, J.~W.}, \au{Kallinikos, N.} \& \au{MacKay, R.~S}} \yr{2020}
  \at{Some mathematics for quasi-symmetry}.  \jt{Journal of Mathematical
  Physics}  \bvol{61}~(9).

\bibitem[Eriksson {\em et~al.\/}(2019)Eriksson, Pearce, Gardner, Turner \&
  Poloczek]{turbo}
{\sc \au{Eriksson, D.}, \au{Pearce, M.}, \au{Gardner, J.}, \au{Turner, R.~D.}
  \& \au{Poloczek, M.}} \yr{2019} Scalable global optimization via local
  {Bayesian} optimization.  \bt{In {\em Advances in Neural Information
  Processing Systems\/}},  \pg{pp. 5496--5507}.

\bibitem[Fuller~Jr(1999)]{fuller1999geometric}
{\sc \au{Fuller~Jr, E.~J.}} \yr{1999} {\em The geometric and topological
  structure of holonomic knots\/}.  \publ{University of Georgia}.

\bibitem[Garren \& Boozer(1991)]{garren1991existence}
{\sc \au{Garren, D.~A.} \& \au{Boozer, A.~H.}} \yr{1991}  \at{Existence of
  quasihelically symmetric stellarators}.  \jt{Physics of Fluids B}
  \bvol{3}~(10),  \pg{2822--2834}.

\bibitem[Giuliani(2024)]{quasr1}
{\sc \au{Giuliani, A.}} \yr{2024}  \at{Direct stellarator coil design using
  global optimization: application to a comprehensive exploration of
  quasi-axisymmetric devices}.  \jt{Journal of Plasma Physics}  \bvol{90}~(3),
  \pg{905900303}.

\bibitem[Giuliani {\em et~al.\/}(2023)Giuliani, Wechsung, Cerfon, Landreman \&
  Stadler]{surfaceopt2}
{\sc \au{Giuliani, A.}, \au{Wechsung, F.}, \au{Cerfon, A.}, \au{Landreman, M.}
  \& \au{Stadler, G.}} \yr{2023}  \at{{Direct stellarator coil optimization for
  nested magnetic surfaces with precise quasi-symmetry}}.  \jt{Physics of
  Plasmas}  \bvol{30}~(4), 042511,  \arxiv{arXiv:
  https://pubs.aip.org/aip/pop/article-pdf/doi/10.1063/5.0129716/16831861/042511\_1\_5.0129716.pdf}.

\bibitem[Giuliani {\em et~al.\/}(2022{\natexlab{{\em a\/}}})Giuliani, Wechsung,
  Cerfon, Stadler \& Landreman]{nae}
{\sc \au{Giuliani, A.}, \au{Wechsung, F.}, \au{Cerfon, A.}, \au{Stadler, G.} \&
  \au{Landreman, M.}} \yr{2022{\natexlab{{\em a\/}}}}  \at{Single-stage
  gradient-based stellarator coil design: Optimization for near-axis
  quasi-symmetry}.  \jt{Journal of Computational Physics}  \bvol{459},
  \pg{111147}.

\bibitem[Giuliani {\em et~al.\/}(2022{\natexlab{{\em b\/}}})Giuliani, Wechsung,
  Stadler, Cerfon \& Landreman]{surfaceopt1}
{\sc \au{Giuliani, A.}, \au{Wechsung, F.}, \au{Stadler, G.}, \au{Cerfon, A.} \&
  \au{Landreman, M.}} \yr{2022{\natexlab{{\em b\/}}}}  \at{Direct computation
  of magnetic surfaces in {B}oozer coordinates and coil optimization for
  quasisymmetry}.  \jt{Journal of Plasma Physics}  \bvol{88}~(4),
  \pg{905880401}.

\bibitem[Helander(2014)]{helander2014theory}
{\sc \au{Helander, P.}} \yr{2014}  \at{Theory of plasma confinement in
  non-axisymmetric magnetic fields}.  \jt{Reports on Progress in Physics}
  \bvol{77}~(8),  \pg{087001}.

\bibitem[Helander \& Simakov(2008)]{PhysRevLett.101.145003}
{\sc \au{Helander, P.} \& \au{Simakov, A.~N.}} \yr{2008}  \at{Intrinsic
  ambipolarity and rotation in stellarators}.  \jt{Phys. Rev. Lett.}
  \bvol{101},  \pg{145003}.

\bibitem[Hindenlang {\em et~al.\/}(2024)Hindenlang, Plunk \&
  Maj]{hindenlang2024generalized}
{\sc \au{Hindenlang, Florian~J}, \au{Plunk, Gabriel~G} \& \au{Maj, Omar}}
  \yr{2024}  \at{A generalized {F}renet frame for computing {MHD} equilibria in
  stellarators}.  \jt{arXiv preprint arXiv:2410.17595} .

\bibitem[Hirshman \& Whitson(1983{\natexlab{{\em a\/}}})]{hirshman1983}
{\sc \au{Hirshman, S.~P.} \& \au{Whitson, J.~C.}} \yr{1983{\natexlab{{\em
  a\/}}}}  \at{Steepest‐descent moment method for three‐dimensional
  magnetohydrodynamic equilibria}.  \jt{The Physics of Fluids}  \bvol{26}~(12),
   \pg{3553--3568}.

\bibitem[Hirshman \& Whitson(1983{\natexlab{{\em b\/}}})]{VMEC}
{\sc \au{Hirshman, S.~P.} \& \au{Whitson, J.~P}} \yr{1983{\natexlab{{\em
  b\/}}}}  \at{Steepest-descent moment method for three-dimensional
  magnetohydrodynamic equilibria.}  \jt{PF}  \bvol{26}~(12),  \pg{3553}.

\bibitem[Jolliffe(1990)]{jolliffe1990principal}
{\sc \au{Jolliffe, I.~T.}} \yr{1990}  \at{Principal component analysis: a
  beginner's guide—{I}. {I}ntroduction and application}.  \jt{Weather}
  \bvol{45}~(10),  \pg{375--382}.

\bibitem[Jorge {\em et~al.\/}(2024)Jorge, Giuliani \&
  Loizu]{jorge2024simplifiedflexiblecoilsstellarators}
{\sc \au{Jorge, R.}, \au{Giuliani, A.} \& \au{Loizu, J.}} \yr{2024}
  \at{Simplified and flexible coils for stellarators using single-stage
  optimization}.  \jt{Physics of Plasmas}  \bvol{31}~(11),  \pg{112501},
  \arxiv{arXiv:
  https://pubs.aip.org/aip/pop/article-pdf/doi/10.1063/5.0226688/20231193/112501\_1\_5.0226688.pdf}.

\bibitem[Jorge {\em et~al.\/}(2023)Jorge, Goodman, Landreman, Rodrigues \&
  Wechsung]{Jorge_2023}
{\sc \au{Jorge, R}, \au{Goodman, A}, \au{Landreman, M}, \au{Rodrigues, J} \&
  \au{Wechsung, F}} \yr{2023}  \at{Single-stage stellarator optimization:
  combining coils with fixed boundary equilibria}.  \jt{Plasma Physics and
  Controlled Fusion}  \bvol{65}~(7),  \pg{074003}.

\bibitem[Kappel {\em et~al.\/}(2024)Kappel, Landreman \&
  Malhotra]{kappel_magnetic_2024}
{\sc \au{Kappel, J.}, \au{Landreman, M.} \& \au{Malhotra, D.}} \yr{2024}
  \at{The magnetic gradient scale length explains why certain plasmas require
  close external magnetic coils}.  \jt{Plasma Physics and Controlled Fusion}
  \bvol{66}~(2),  \pg{025018}.

\bibitem[Kaptanoglu {\em et~al.\/}(2024)Kaptanoglu, Langlois \&
  Landreman]{KAPTANOGLU2024116504}
{\sc \au{Kaptanoglu, Alan~A.}, \au{Langlois, Gabriel~P.} \& \au{Landreman,
  Matt}} \yr{2024}  \at{Topology optimization for inverse magnetostatics as
  sparse regression: Application to electromagnetic coils for stellarators}.
  \jt{Computer Methods in Applied Mechanics and Engineering}  \bvol{418},
  \pg{116504}.

\bibitem[Kaptanoglu {\em et~al.\/}(2025)Kaptanoglu, Wiedman, Halpern, Hurwitz,
  Paul \& Landreman]{Kaptanoglu_2025}
{\sc \au{Kaptanoglu, Alan~A.}, \au{Wiedman, Alexander}, \au{Halpern, Jacob},
  \au{Hurwitz, Siena}, \au{Paul, Elizabeth~J.} \& \au{Landreman, Matt}}
  \yr{2025}  \at{Reactor-scale stellarators with force and torque minimized
  dipole coils}.  \jt{Nuclear Fusion}  \bvol{65}~(4),  \pg{046029}.

\bibitem[Kobak \& Berens(2019)]{kobak2019art}
{\sc \au{Kobak, D.} \& \au{Berens, P.}} \yr{2019}  \at{The art of using t-{SNE}
  for single-cell transcriptomics}.  \jt{Nature communications}  \bvol{10}~(1),
   \pg{5416}.

\bibitem[Kruskal(1964)]{kruskal1964nonmetric}
{\sc \au{Kruskal, J.~B.}} \yr{1964}  \at{Nonmetric multidimensional scaling: a
  numerical method}.  \jt{Psychometrika}  \bvol{29}~(2),  \pg{115--129}.

\bibitem[Landreman(2017)]{Landreman_2017}
{\sc \au{Landreman, Matt}} \yr{2017}  \at{An improved current potential method
  for fast computation of stellarator coil shapes}.  \jt{Nuclear Fusion}
  \bvol{57}~(4),  \pg{046003}.

\bibitem[Landreman(2019)]{Landreman_2019}
{\sc \au{Landreman, M.}} \yr{2019}  \at{Optimized quasisymmetric stellarators
  are consistent with the {G}arren–{B}oozer construction}.  \jt{Plasma
  Physics and Controlled Fusion}  \bvol{61}~(7),  \pg{075001}.

\bibitem[Landreman(2022)]{Landreman_2022}
{\sc \au{Landreman, M.}} \yr{2022}  \at{Mapping the space of quasisymmetric
  stellarators using optimized near-axis expansion}.  \jt{Journal of Plasma
  Physics}  \bvol{88}~(6),  \pg{905880616}.

\bibitem[Landreman {\em et~al.\/}(2024)Landreman, Jorge, Rodriguez \&
  Dudt]{pyqsc}
{\sc \au{Landreman, M.}, \au{Jorge, R.}, \au{Rodriguez, E.} \& \au{Dudt, D.}}
  \yr{2024} {\em landreman/pyQSC\/}.  \publ{GitHub repository}.

\bibitem[Landreman {\em et~al.\/}(2021)Landreman, Medasani, Wechsung, Giuliani,
  Jorge \& Zhu]{landreman2021simsopt}
{\sc \au{Landreman, M.}, \au{Medasani, B.}, \au{Wechsung, F.}, \au{Giuliani,
  A.}, \au{Jorge, R.} \& \au{Zhu, C.}} \yr{2021}  \at{{SIMSOPT}: a flexible
  framework for stellarator optimization}.  \jt{Journal of Open Source
  Software}  \bvol{6}~(65),  \pg{3525}.

\bibitem[Landreman \& Paul(2022)]{lp}
{\sc \au{Landreman, M.} \& \au{Paul, E.}} \yr{2022}  \at{Magnetic fields with
  precise quasisymmetry for plasma confinement}.  \jt{Phys. Rev. Lett.}
  \bvol{128},  \pg{035001}.

\bibitem[Landreman \& Sengupta(2019)]{Landreman_Sengupta_2019}
{\sc \au{Landreman, M.} \& \au{Sengupta, W.}} \yr{2019}  \at{Constructing
  stellarators with quasisymmetry to high order}.  \jt{Journal of Plasma
  Physics}  \bvol{85}~(6),  \pg{815850601}.

\bibitem[Landreman {\em et~al.\/}(2019)Landreman, Sengupta \&
  Plunk]{landreman2019direct}
{\sc \au{Landreman, M.}, \au{Sengupta, W.} \& \au{Plunk, G.~G.}} \yr{2019}
  \at{Direct construction of optimized stellarator shapes. {P}art 2.
  {N}umerical quasisymmetric solutions}.  \jt{Journal of Plasma Physics}
  \bvol{85}~(1),  \pg{905850103}.

\bibitem[Lee \& Verleysen(2009)]{lee2009quality}
{\sc \au{Lee, J.~A.} \& \au{Verleysen, M.}} \yr{2009}  \at{Quality assessment
  of dimensionality reduction: Rank-based criteria}.  \jt{Neurocomputing}
  \bvol{72}~(7-9),  \pg{1431--1443}.

\bibitem[Van~der Maaten \& Hinton(2008)]{van2008visualizing}
{\sc \au{Van~der Maaten, L.} \& \au{Hinton, G.}} \yr{2008}  \at{Visualizing
  data using t-{SNE}.}  \jt{Journal of machine learning research}
  \bvol{9}~(11).

\bibitem[Moffatt \& Ricca(1992)]{moffatt1992helicity}
{\sc \au{Moffatt, H.~K.} \& \au{Ricca, R.~L.}} \yr{1992}  \at{Helicity and the
  c{\u{a}}lug{\u{a}}reanu invariant}.  \jt{Proceedings of the Royal Society of
  London. Series A: Mathematical and Physical Sciences}  \bvol{439}~(1906),
  \pg{411--429}.

\bibitem[Najmabadi {\em et~al.\/}(2008)Najmabadi, Raffray, Abdel-Khalik,
  Bromberg, Crosatti, El-Guebaly, Garabedian, Grossman, Henderson, Ibrahim {\em
  et~al.\/}]{najmabadi2008aries}
{\sc \au{Najmabadi, F.}, \au{Raffray, A.~R.}, \au{Abdel-Khalik, S.~I.},
  \au{Bromberg, L.}, \au{Crosatti, L.}, \au{El-Guebaly, L.}, \au{Garabedian,
  P.~R.}, \au{Grossman, A.~A.}, \au{Henderson, D.}, \au{Ibrahim, A.} \&
  \au{others}} \yr{2008}  \at{The {ARIES-CS} compact stellarator fusion power
  plant}.  \jt{Fusion Science and Technology}  \bvol{54}~(3),  \pg{655--672}.

\bibitem[Nies {\em et~al.\/}(2024)Nies, Paul, Panici, Hudson \&
  Bhattacharjee]{nies2024exploration}
{\sc \au{Nies, R.}, \au{Paul, E.~J.}, \au{Panici, D.}, \au{Hudson, S.~R.} \&
  \au{Bhattacharjee, A.}} \yr{2024}  \at{Exploration of the parameter space of
  quasisymmetric stellarator vacuum fields through adjoint optimisation}.
  \jt{arXiv preprint arXiv:2404.02240} .

\bibitem[Nührenberg \& Zille(1988)]{NUHRENBERG1988113}
{\sc \au{Nührenberg, J.} \& \au{Zille, R.}} \yr{1988}  \at{Quasi-helically
  symmetric toroidal stellarators}.  \jt{Physics Letters A}  \bvol{129}~(2),
  \pg{113--117}.

\bibitem[Oberti \& Ricca(2016)]{oberti2016torus}
{\sc \au{Oberti, C.} \& \au{Ricca, R.~L.}} \yr{2016}  \at{On torus knots and
  unknots}.  \jt{Journal of Knot Theory and Its Ramifications}  \bvol{25}~(06),
   \pg{1650036}.

\bibitem[Panici {\em et~al.\/}(2023)Panici, Conlin, Dudt, Unalmis \&
  Kolemen]{Panici_Conlin_Dudt_Unalmis_Kolemen_2023}
{\sc \au{Panici, D.}, \au{Conlin, R.}, \au{Dudt, D.W.}, \au{Unalmis, K.} \&
  \au{Kolemen, E.}} \yr{2023}  \at{The {DESC} stellarator code suite. {P}art 1.
  {Q}uick and accurate equilibria computations}.  \jt{Journal of Plasma
  Physics}  \bvol{89}~(3),  \pg{955890303}.

\bibitem[Paul {\em et~al.\/}(2022)Paul, Bhattacharjee, Landreman, Alex, Velasco
  \& Nies]{paul2022energetic}
{\sc \au{Paul, E.~J.}, \au{Bhattacharjee, A.}, \au{Landreman, M.}, \au{Alex,
  D.}, \au{Velasco, J.~L.} \& \au{Nies, R.}} \yr{2022}  \at{Energetic particle
  loss mechanisms in reactor-scale equilibria close to quasisymmetry}.
  \jt{Nuclear Fusion}  \bvol{62}~(12),  \pg{126054}.

\bibitem[Rodriguez {\em et~al.\/}(2020)Rodriguez, Helander \&
  Bhattacharjee]{rodriguez2020necessary}
{\sc \au{Rodriguez, E.}, \au{Helander, P.} \& \au{Bhattacharjee, A.}} \yr{2020}
   \at{Necessary and sufficient conditions for quasisymmetry}.  \jt{Physics of
  Plasmas}  \bvol{27}~(6).

\bibitem[Rodriguez {\em et~al.\/}(2022{\natexlab{{\em a\/}}})Rodriguez, Paul \&
  Bhattacharjee]{rodriguez2022measures}
{\sc \au{Rodriguez, E.}, \au{Paul, E.~J.} \& \au{Bhattacharjee, A.}}
  \yr{2022{\natexlab{{\em a\/}}}}  \at{Measures of quasisymmetry for
  stellarators}.  \jt{Journal of Plasma Physics}  \bvol{88}~(1),
  \pg{905880109}.

\bibitem[Rodriguez {\em et~al.\/}(2022{\natexlab{{\em b\/}}})Rodriguez,
  Sengupta \& Bhattacharjee]{rodriguez2022phases}
{\sc \au{Rodriguez, E.}, \au{Sengupta, W.} \& \au{Bhattacharjee, A.}}
  \yr{2022{\natexlab{{\em b\/}}}}  \at{Phases and phase-transitions in
  quasisymmetric configuration space}.  \jt{Plasma Physics and Controlled
  Fusion}  \bvol{64}~(10),  \pg{105006}.

\bibitem[Rodr{\'\i}guez {\em et~al.\/}(2023)Rodr{\'\i}guez, Sengupta \&
  Bhattacharjee]{rodriguez2023constructing}
{\sc \au{Rodr{\'\i}guez, E.}, \au{Sengupta, W.} \& \au{Bhattacharjee, A.}}
  \yr{2023}  \at{Constructing the space of quasisymmetric stellarators through
  near-axis expansion}.  \jt{Plasma Physics and Controlled Fusion}
  \bvol{65}~(9),  \pg{095004}.

\bibitem[Sanchez {\em et~al.\/}(2000)Sanchez, Hirshman, Ware, Berry \&
  Spong]{sanchez2000ballooning}
{\sc \au{Sanchez, R.}, \au{Hirshman, S.~P.}, \au{Ware, A.~S.}, \au{Berry,
  L.~A.} \& \au{Spong, D.~A.}} \yr{2000}  \at{Ballooning stability optimization
  of low-aspect-ratio stellarators}.  \jt{Plasma physics and controlled fusion}
   \bvol{42}~(6),  \pg{641}.

\bibitem[Sengupta {\em et~al.\/}(2023)Sengupta, Nikulsin, Buller, Madan, Paul,
  Nies, Kaptanoglu, Hudson \& Bhattacharjee]{sengupta2023periodic}
{\sc \au{Sengupta, W}, \au{Nikulsin, N}, \au{Buller, S}, \au{Madan, R},
  \au{Paul, EJ}, \au{Nies, R}, \au{Kaptanoglu, AA}, \au{Hudson, SR} \&
  \au{Bhattacharjee, A}} \yr{2023}  \at{Periodic {K}orteweg-de {V}ries soliton
  potentials generate magnetic field strength with excellent quasisymmetry}.
  \jt{arXiv preprint arXiv:2302.13924} .

\bibitem[Spitzer(1958)]{spitzer1958stellarator}
{\sc \au{Spitzer, L.}} \yr{1958}  \at{The stellarator concept}.  \jt{Physics of
  Fluids}  \bvol{1}~(4),  \pg{253}.

\bibitem[Tenenbaum {\em et~al.\/}(2000)Tenenbaum, Silva \&
  Langford]{tenenbaum2000global}
{\sc \au{Tenenbaum, J.~B.}, \au{Silva, V.~de} \& \au{Langford, J.~C.}}
  \yr{2000}  \at{A global geometric framework for nonlinear dimensionality
  reduction}.  \jt{science}  \bvol{290}~(5500),  \pg{2319--2323}.

\bibitem[Trefethen \& Weideman(2014)]{trefethen2014exponentially}
{\sc \au{Trefethen, L.~N.} \& \au{Weideman, J. A.~C.}} \yr{2014}  \at{The
  exponentially convergent trapezoidal rule}.  \jt{SIAM review}  \bvol{56}~(3),
   \pg{385--458}.

\bibitem[Wechsung {\em et~al.\/}(2022{\natexlab{{\em a\/}}})Wechsung, Giuliani,
  Landreman, Cerfon \& Stadler]{Wechsung_2022}
{\sc \au{Wechsung, Florian}, \au{Giuliani, Andrew}, \au{Landreman, Matt},
  \au{Cerfon, Antoine} \& \au{Stadler, Georg}} \yr{2022{\natexlab{{\em a\/}}}}
  \at{Single-stage gradient-based stellarator coil design: stochastic
  optimization}.  \jt{Nuclear Fusion}  \bvol{62}~(7),  \pg{076034}.

\bibitem[Wechsung {\em et~al.\/}(2022{\natexlab{{\em b\/}}})Wechsung,
  Landreman, Giuliani, Cerfon \& Stadler]{wechsung2022precise}
{\sc \au{Wechsung, F.}, \au{Landreman, M.}, \au{Giuliani, A.}, \au{Cerfon, A.}
  \& \au{Stadler, G.}} \yr{2022{\natexlab{{\em b\/}}}}  \at{Precise stellarator
  quasi-symmetry can be achieved with electromagnetic coils}.  \jt{Proceedings
  of the National Academy of Sciences}  \bvol{119}~(13),  \pg{e2202084119}.

\bibitem[Wiedman {\em et~al.\/}(2024)Wiedman, Buller \&
  Landreman]{Wiedman_Buller_Landreman_2024}
{\sc \au{Wiedman, A.}, \au{Buller, S.} \& \au{Landreman, M.}} \yr{2024}
  \at{Coil optimization for quasi-helically symmetric stellarator
  configurations}.  \jt{Journal of Plasma Physics}  \bvol{90}~(3),
  \pg{905900307}.

\bibitem[Zhu {\em et~al.\/}(2017)Zhu, Hudson, Song \& Wan]{Zhu_2018}
{\sc \au{Zhu, Caoxiang}, \au{Hudson, Stuart~R.}, \au{Song, Yuntao} \& \au{Wan,
  Yuanxi}} \yr{2017}  \at{New method to design stellarator coils without the
  winding surface}.  \jt{Nuclear Fusion}  \bvol{58}~(1),  \pg{016008}.

\end{thebibliography}

\end{document}